\documentclass{aa}  
\usepackage{amsmath,amssymb}
\usepackage{graphicx,float,subfigure}
\usepackage{txfonts}
\usepackage{natbib}
\bibpunct{(}{)}{;}{a}{}{,}
\newcommand{\vect}[1]{\ensuremath{\boldsymbol{#1}}}
\newcommand{\dd}{\ensuremath{\mathrm{d}}}
\newcommand{\DD}{\ensuremath{\mathrm{D}}}
\newcommand{\unit}[1]{\ensuremath{\,\mathrm{pc}}}

\begin{document}

\title{Numerical simulations of compressively driven interstellar turbulence: I. Isothermal gas}
\titlerunning{Compressively driven turbulence in isothermal gas}

\author{Wolfram Schmidt\inst{1}, Christoph Federrath\inst{2}, Markus Hupp\inst{1},\\
        Sebastian Kern\inst{1} and Jens C. Niemeyer\inst{1}}
\authorrunning{W. Schmidt et al.}

\institute{Lehrstuhl f\"{u}r Astronomie, Institut f\"{u}r Theoretische Physik und Astrophysik,
  Universit\"{a}t W\"{u}rzburg, Am Hubland, D-97074 W\"{u}rzburg, Germany \and
  Institut f\"{u}r Theoretische Astrophysik, Universit\"{a}t Heidelberg, Albert-Ueberle-Str.~2, D-69120 Heidelberg, Germany}

\date{Received / Accepted}

\abstract
{Supersonic turbulence is ubiquitous in the interstellar medium and plays an important role in contemporary star formation.}
{To perform a high-resolution numerical simulation of supersonic isothermal turbulence driven by
compressive large-scale forcing and to analyse various statistical properties.}
{The compressible Euler equations with an external stochastic force field dominated by rotation-free modes are solved with the piecewise parabolic method. Both a static grid and adaptive mesh refinement is used with an effective resolution $N=768^{3}$.}
{After a transient phase dominated by shocks, turbulence evolves into a steady state with
root mean square Mach number $\approx 2.2\ldots 2.5$, in which cloud-like structures of over-dense gas are surrounded by highly rarefied gas. The index of the turbulence energy spectrum function $\beta\approx 2.0$ in the shock-dominated phase. As the flow approaches statistical equilibrium, the spectrum flattens, with $\beta\approx 1.9$. For the scaling exponent of the root mean square velocity fluctuation, we obtain $\gamma\approx 0.43$ from the velocity structure functions of second order. These results are well within the range of observed scaling properties for the velocity dispersion in molecular clouds. Calculating structure functions of order $p=1,\ldots,5$, we find for \emph{all} scaling exponents significant deviations from the Kolmogorov-Burgers model proposed by Boldyrev. Our results are very well described by a general log-Poisson model with a higher degree of intermittency,
which implies an influence of the forcing on the scaling properties. The spectral index of the quadratic logarithmic density fluctuation $\beta_{\delta}\approx 1.8$. Contrary to previous numerical results for isothermal turbulence, we obtain a skewed probability density function of the mass density fluctuations that is not consistent with log-normal statistics and entails a substantially higher fraction of mass in the density peaks than implied by the Padoan-Nordlund relation between the variance of the density fluctuations and the Mach number.}
{Even putting aside further complexity due to magnetic fields, gravity or thermal processes, we question the notion that Larson-type relations are a consequence of universal supersonic turbulence scaling. For a genuine understanding, it seems necessary to account for the production mechanism of turbulence in the ISM.}
\keywords{Hydrodynamics - Turbulence - Methods: numerical - ISM: kinematics and dynamics}

\maketitle

\section{Introduction}

Today it is agreed that supersonic turbulence occurs over a large range of length scales in the interstellar medium (ISM) and, particularly, within molecular clouds \citep{ElmeScal04}. This is inferred from different observations of molecular line broadening which indicate, firstly, a velocity dispersion comparable to or greater than the speed of sound and, secondly, a power-law dependence of the velocity dispersion, $\Delta v\propto \ell^{\,\alpha}$, in the range of length scales $0.1\,\mathrm{pc}\lesssim\ell\lesssim10\,\mathrm{pc}$ of the associated structures \citep{Larson81,FalPu92,MieBal94,BruntHey02b}. This has been interpreted as a manifestation of the universality of interstellar turbulence \citep{HeyBrunt04} and bears important consequences on the theory of turbulence-regulated star formation, where Larson-type relations for the velocity dispersion are explained as a direct consequence of the self-similarity of the turbulence cascade, with possible modifications of the scaling exponent due to magnetic fields, self-gravity or thermal processes \citep{PadoanNordlund02,LowKless04,BallKless07,KeeOst07}.

The question of the origin of supersonic interstellar turbulence remains largely open \citep{ElmeScal04,BallKless07}. For the numerical simulation of interstellar turbulence, there is also the difficulty that energy injection into the interstellar medium and processes leading to star formation may occur on vastly disparate scales, which cannot be encompassed by a single numerical simulation. This is why two major approaches to the problem have emerged: Either the generation of turbulence by instabilities at relatively large scales is computed or statistically isotropic turbulence is simulated in a periodic box as a simple model for the small-scale regime. Typical examples for the first approach are the simulations of colliding flows by  \citet{HeitSly06}, \citet{VazGom07} and \citet{HenneAud07}. Recent simulations of driven turbulence in a periodic box were presented by \citet{PadNord07} and \citet{KritNor07}. While colliding flow simulations follow the idea of controlling the flow evolution in the computational domain via initial and boundary conditions only, the meaning of turbulence simulations with a driving force are implicitly based on the notion of an \emph{upper} numerical cutoff scale. This means that one imagines turbulence being produced on scales larger than the size of the computational box and the driving force is intended to mimic the interactions with the larger scales in a heuristic fashion. Indeed, decomposing the hydrodynamic equations into a large-scale part and a small-scale part entails coupling via an energy flux that, in the most simplified manner, can be modelled as a random volume force acting upon the small-scale part \cite{Frisch}. Such a volume force can produce shear or compression by exciting solenoidal (divergence-free) or dilatational (rotation-free) modes of the velocity field.

In this article, we follow a hybrid approach. A statistically isotropic stochastic force field that excites mostly dilatational modes is applied in our simulations. Thereby, converging supersonic flows are produced randomly at large scales. From the collisions of these flows in combination with a small solenoidal component of the stochastic force, turbulence is generated and the flow evolves towards a steady state, in which energy injection due to the forcing is balanced by numerical dissipation. We refer to this scenario as compressively driven turbulence. Apart from the dominance of the rotation-free component, our method differs in another aspect form the forcing applied, for instance, by \citet{KritNor07} and in many earlier simulations. Whereas they imprint a certain large-scale structure by a steady random force that is derived from some initial velocity field and merely replenishes the energy of the flow, we start with isothermal gas of uniform density at rest and let turbulence develop dynamically. The spatiotemporal correlation properties of the stochastic force field guarantee that large-scale structures evolve on a time scale that is in accordance with their characteristic velocity and size. Moreover, a statistically isotropic state is approached regardless of the initial condition, because temporal correlations decay exponentially. There is a common point of view though that the mode of energy injection in driven turbulence simulations is not important for statistical properties of the resulting flow \citep{Boldyrev+02}. The statistical analysis of our simulation data, however, cast doubt on his claim. Our results indicate that the dynamics of compressible turbulence at length scales significantly smaller than the energy-containing scale is affected by the forcing. For this reason,
the notion of fully developed turbulence--i.~e., turbulence that is not affected by imposed constraints such as boundary conditions, forcing or viscosity--is possibly meaningless in the highly compressible regime. Of course, we cannot rule out that universal statistics might be asymptotically approached toward length scales smaller than the resolution of our simulation. We note, however, that universality of the small-scale dynamics has been questioned even in the case of incompressible turbulence \citep{AlexaMin05,CasGual07}.

The isothermal approximation is expected to apply to the interiors of molecular clouds, where
the temperature of the cold phase of the interstellar gas is close to its minimum \citep{BallKless07}. 
In molecular clouds, there are prominent magnetic fields. Although they have indisputably a large impact on fragmentation properties \citep{HenneTeys08}, it appears that scaling exponents are not greatly altered by magnetic fields, at least in the super-Alfv\'{e}nic case \citep{Boldyrev+02,PadJim04,PadNord07}. Consequently, our analysis of the velocity statistics of compressively
driven turbulence has potential relevance to molecular cloud turbulence.  At larger scales in the ISM, various heating and cooling processes bring about thermal instabilities. These are included in the colliding flow simulations by \citet{HeitSly06}, \citet{VazGom07} and \citet{HenneAud07}. We will consider driven turbulence in thermally bistable gas in a follow-up paper. In the future, we also plan to include self-gravity and magnetic fields. By successively including more complex physics, we expect to achieve a thorough understanding of the effects stemming from certain physical processes. The numerical treatment of processes such as cooling or self-gravity poses the problem of covering a wide range of length scales, at least in some regions of the flow. Using grid-based codes, a possible solution is adaptive mesh refinement (AMR). The simulation of turbulence with AMR was pioneered by \citet{KritNor06}. Recently, an AMR simulations of self-gravitating turbulence have been presented \citep{OffKrum07}. As a preliminary study, we also performed an AMR simulation, with the rate of compression (i.~e., the rate of change of the negative divergence) and the vorticity modulus as control variables for refinement. Velocity and density statistics are well reproduced in comparison to the simulation with a static grid, but, using only one level of refinement, we have not achieved a significant gain in terms of performance in the case of isothermal supersonic turbulence. The same difficulty was encountered by \citep{KritNor07} for more than twice the resolution we used. 

The article is organised as follows. Section~\ref{sc:numerics} explains numerical techniques, in particular, the stochastic forcing and the method of refinement employed in the AMR simulation. In Section~\ref{sc:results}, we begin our presentation of simulation results with global statistics and interpret some properties via three-dimensional renderings of the mass density and the vorticity in the stage of turbulence production. In the following, we analyse the probability density functions of the mass density and the vorticity. Next we turn our attention to the spectra of the turbulence energy and the density fluctuations. Finally, we derive scaling exponents from velocity structure functions. The results are related to each other and compared to observational results, theoretical predictions as well as 
results from other numerical simulations in Section~\ref{sc:discussion}. In the conclusion, we discuss possible implications on molecular cloud turbulence and star formation and give an outlook to future work.

\section{Numerical techniques}
\label{sc:numerics}

For the simulations presented in this article, we employed the open
source code Enzo \citep{SheaBry04} in order to solve the compressible Euler equations
by means of the piecewise-parabolic method (PPM) of
\cite{ColWood84}. The Euler equations including an external force
density $\rho\vect{f}$ can be written as follows:
\begin{align}
  \label{eq:mass}
  &\frac{\DD}{\DD t}\rho = -\rho\vect{\nabla}\cdot\vect{u}, \\
  \label{eq:vel}
  \rho & \frac{\DD}{\DD t}\vect{u} =
  -\vect{\nabla}P + \rho\vect{f}, \\
  \label{eq:energy_tot}
  \rho &\frac{\DD}{\DD t}e + \vect{\nabla}\cdot P\vect{u} =
  \rho\vect{f}\cdot\vect{u}.
\end{align}
The primitive variables are the mass density $\rho$, the velocity
$\vect{u}$ and the specific total energy $e$ of the fluid. The Lagrangian time derivative is defined by
\begin{equation}
    \frac{\DD}{\DD t} =
    \frac{\partial}{\partial t} + \vect{u}\cdot\vect{\nabla}.
\end{equation}
The total energy per unit mass is given by
\begin{equation}
    e = \frac{1}{2}u^{2} + \frac{P}{(\gamma-1)\rho},
\end{equation}
where $\gamma$ is the adiabatic exponent and the pressure $P$ is
related to the mass density and the temperature $\mathcal{T}$ via the ideal
gas law:
\begin{equation}
    P=\frac{\rho k_{\mathrm{B}}\mathcal{T}}{\mu m_{\mathrm{H}}}.
\end{equation}
The constants $k_{\mathrm{B}}$, $\mu$ and $m_{\mathrm{H}}$
denote the Boltzmann constant, the mean molecular
weight and the mass of the hydrogen atom, respectively .

The specific force $\vect{f}$ in equations~(\ref{eq:vel}) and~(\ref{eq:energy_tot}) smoothly accelerates the fluid at large scales. Using a periodic domain, we adopt a generalisation of the Ornstein-Uhlenbeck process for the Fourier modes $\hat{\vect{f}}(\vect{k},t)$ of the acceleration field \citep{EswaPope88,SchmHille06}:
\begin{equation}
\label{eq:forcing}
        \dd\hat{\vect{f}}(\vect{k},t) = g_{\zeta}\left[
        -\hat{\vect{f}}(\vect{k},t)\frac{\dd t}{T} +
        \frac{V}{T}\left(\frac{2\sigma^{2}(\vect{k})}{T}\right)^{1/2}
            \vect{P}_{\zeta}(\vect{k})\cdot\dd\vect{\mathcal{W}}_{t}\right].
\end{equation}
In this stochastic differential equation, the Wiener process
$\vect{\mathcal{W}}_{t}$ generates Gaussian random vector deviates. We define the characteristic forcing wavenumber $k_{0}:=2\pi\alpha/X$ for a cubic domain of size $X$. Ideally, one would prefer to have $\alpha\ll 1$ a in order to minimize the effect of periodic boundary
conditions \citep{Davidson}. However, this would constrain the dynamical range by far too much for a
grid size that is computationally feasible. For this reason, we set $\alpha=2$ so that the wavelength
of the force field is about half of the domain size and there are $\sim 10$ energy-containing structures ("large eddies"). To allow for a small spread of forcing wave numbers, we use the parabolic
weighing function $\sigma(\vect{k})\propto k^{2}(2 k_{0}-k)^{2}H(k-2k_{0})$,
where $k=|\vect{k}|$ and $H$ is the Heaviside step function, selects non-zero modes for $|\vect{k}|\in]0,2k_{0}[$. The autocorrelation time scale $T$ of the forcing is identified with the large-eddy turn-over time, i.~e., $T=L/V$, where $V$ is the characteristic velocity of the flow in the stationary regime and $L=X/\alpha=2\pi/k_{0}$ is the integral length. The stochastic diffusion term in equation~(\ref{eq:forcing}) ensures that the resulting force field becomes statistically isotropic in physical space, while any anisotropic initial condition decays exponentially by virtue of the first term
on the right-hand side. This is not guaranteed  with the method of steady random forcing, where
isotropy must be prepared to high precision in the initial velocity field.

For purely solenoidal forcing, the random deviates generated by the Wiener process are projected
perpendicular to the associated wave vectors $\vect{k}$ by means of the projection operator
\begin{equation}
  \label{eq:proj_op}
  (P_{ij})_{\zeta}(\vect{k}) =
  \zeta P_{ij}^{\perp}(\vect{k}) +
  (1-\zeta)P_{ij}^{\parallel}(\vect{k}) =
  \zeta\delta_{ij} + (1-2\zeta)\frac{k_{i}k_{j}}{k^{2}}.
\end{equation}
with $\zeta=1$. In the case $\zeta=0$, on the other hand, projection
parallel to the wave vectors results in a dilatational forcing field. Due to the normalisation factor
\begin{equation}
    g_{\zeta}=\frac{3}{\sqrt{1 - 2\zeta + 3\zeta^{2}}}
\end{equation}
in equation~(\ref{eq:forcing}), the root mean square acceleration
$f_{\mathrm{rms}}=3V^{2}/L$ is independent of $\zeta$. We chose
$V=5c_{0}/\sqrt{3}$, where $c_{0}$ is the initial speed of sound.
For the projection operator~(\ref{eq:proj_op}), we set $\zeta=0.1$. Thus, the forcing in
our simulations is constructed as superposition of a compressive 
component that produces supersonic flow with Mach number roughly
given by $\mathrm{Ma}=V/c_{0}\approx 2.9$ and a subdominant solenoidal
component. In consequence, there are two mechanisms of vorticity
generation: Firstly, large-scale eddies are directly generated by the solenoidal
forcing component. Secondly, the collision of shock fronts produced by the compressive
component of the driving force results in vorticity generation due to small-scale instabilities
\citep{WaldFol98}. In contrast to \citet{KritNor07}, the second mechanism
is more important in our simulations.

In the limiting case that all heat produced by kinetic energy dissipation is
removed from the system immediately, the gas
evolves isothermally. Nearly isothermal turbulence can be
simulated by setting the adiabatic exponent $\gamma$ to a value slightly greater than unity. 
Then $P\simeq\rho c_{\mathrm{s}}^{2}$, where $c_{\mathrm{s}}$ is approximately equal
to the constant isothermal speed of sound. For $\gamma=1.01$ and $\mathrm{Ma}^{2}\lesssim 10$, as in our simulation, the specific kinetic energy $e_{\mathrm{kin}}\lesssim
10c_{\mathrm{s}}^{2}$ and the internal energy density
$e_{\mathrm{int}}=P/(\gamma-1)\rho\sim 100c_{\mathrm{s}}^{2}$. Thus,
the dissipation of kinetic energy results in a secular change of
the internal energy. In this sense, the gas is nearly isothermal. As pointed out by \citet{KritNor07},
the mean helicity of isothermal turbulence should be conserved. In general, the
velocity increments produced by random forces are not free of helicity locally but average
out globally. The helicity of a dilatational force, however, is identically zero.

The main simulation run on a static grid of $N=768^{3}$ cells, while
a root grid of $N_{0}=196^{3}$ and one level of refinement with an increase of resolution by a
factor $4$ was applied in an AMR simulation. The effective resolution,
$N_{1,\mathrm{eff}}=768^{3}$, matches the resolution of the static
grid run. An essential requirement for computing turbulent flow with AMR is the sensitivity
of the refinement criteria on small-scale properties of the flow. For this reason, gradients of
the velocity, the density or other fields are commonly used as control variables for
refinement \citep{KritNor07}. In this article, we focus on \emph{structural invariants}, i.~e., scalars derived from the velocity gradient. A choice that suggests itself for tracking down turbulent eddies is the enstrophy.
Thus, we define the control variable $h_{1}=\frac{1}{2}\omega^{2}$, 
where the vorticity $\vect{\omega}=\vect{\nabla}\times\vect{u}$. 
In supersonic turbulence, the steepening of density gradients is associated
with rising gas compression. This is indicated by a positive rate of convergence, i.~e., the
negative time derivative of the divergence $d=\vect{\nabla}\cdot\vect{u}$.
Letting the operator $\vect{\nabla}\cdot$ act upon
equation~(\ref{eq:vel}), it can be shown that the rate of
convergence is given by
\begin{equation}
    \label{eq:div}
    -\frac{\mathrm{D}}{\mathrm{D}t}d =
    \frac{1}{2}\left(|S|^{2}-\omega^{2}\right)
   +\vect{\nabla}\cdot\left(\frac{1}{\rho}\vect{\nabla}P-\vect{f}\right).
\end{equation}
In the isothermal limit, this equation becomes
\begin{equation}
   \label{eq:div_isoth}
    -\frac{\mathrm{D}}{\mathrm{D}t}d =
    \frac{1}{2}\left(|S|^{2}-\omega^{2}\right) + c_{\mathrm{0}}^{2}\nabla^{2}\delta - \vect{\nabla}\cdot\vect{f}
\end{equation}
where $\delta=\ln(\rho/\rho_{0})$. Dropping the large-scale forcing term on the right hand side, which is not supposed to trigger refinement, we define the control variable
\begin{equation}
    h_{2} \simeq
    \frac{1}{2}\left(|S|^{2}-\omega^{2}\right)
    + c_{0}^{2}\nabla^{2}\delta.
\end{equation}
Note that $h_{2}\simeq 0$ in the incompressible limit. In more general applications such as turbulence in non-isothermal or self-gravitating gas, the corresponding effects can be singled out as additional terms in $h_{2}$.

\begin{table*}[t]
  \begin{center}
    \begin{tabular}{lrcccrrrrcc}
      \hline
      $t/T$  &  $\rho_{\mathrm{max}}$ & $\sigma(\rho$) &
      $P(v/c_{\mathrm{s}}>1)$  &  $\mathcal{M}_\mathrm{rms}$  &  
      $\omega_{\mathrm{rms}}$  &  $\langle\rho\omega^{2}\rangle^{1/2}$  & 
      $\lambda/\Delta$  &  $\lambda_{\mathrm{mw}}/\Delta$  & 
      $r_{\mathrm{cs}}$  & $r_{\mathrm{csmw}}$ \\ \hline\hline
      $0.15$    &       2.3 &  0.21& $0.44$ &  $1.04$ &      2.5 &     2.6 & 707.1 & 687.6 & 0.97 & 0.98\\       \hline
      $0.29$    &     76.9 & 1.05 & $0.86$ & $1.92$ &      5.4 &     7.9 & 656.4 & 382.5 & 0.97 & 0.97\\       \hline
      $0.44$    & 1335.0 & 3.72 & $0.93$ & $2.61$ &   28.8 &   91.9 & 154.9 &   37.5 & 0.92 & 0.77\\       \hline
      $0.58$    &   495.4 & 3.71 & $0.95$ &  $2.92$ &   86.1 & 178.1 &   57.9 &   20.8 & 0.87 & 0.56\\       \hline
      $0.73$    &   336.0 & 3.34 & $0.96$ &  $2.80$ & 137.9 & 234.4 &   34.8 &   16.4 & 0.78 & 0.46\\       \hline
      $0.87$    &   441.1 & 2.95 & $0.95$ &  $2.63$ & 168.5 & 264.9 &   27.0 &   13.6 & 0.67 & 0.41\\       \hline
      $1.02$    &   179.4 & 2.46 & $0.93$ &  $2.49$ & 187.3 & 259.4 &   23.1 &   13.3 & 0.58 & 0.37\\       \hline 
      $2.03$    &   128.8 & 2.65 & $0.93$ &  $2.50$ & 201.0 & 297.0 &   21.7 &   12.1 & 0.51 & 0.29\\       \hline
      $3.05$    &   186.3 & 2.24 & $0.92$ &  $2.44$ & 204.4 & 285.2 &   21.3 &   12.8 & 0.47 & 0.31\\       \hline
      $3.92$    &   259.7 & 2.31 & $0.92$ &  $2.35$ & 207.6 & 294.7 &   20.5 &   13.0 & 0.47 & 0.29\\       \hline
      $5.08$    &   268.4 & 2.99 & $0.91$ &  $2.31$ & 208.8 & 300.3 &   20.2 &   11.5 & 0.48 & 0.27\\       \hline
      $5.95$    &   172.4 & 2.24 & $0.91$ &  $2.23$ & 202.1 & 285.7 &   20.5 &   12.4 & 0.47 & 0.28\\       \hline
      $9.14$    &   383.0 & 3.05 & $0.91$ &  $2.20$ & 210.5 & 312.1 &   20.2 &   12.0 & 0.47 & 0.29\\       \hline
    \end{tabular}
    \caption{Global statistics at various instants of time. From left to right, the columns
    contain the normalised time, peak density, standard deviation of the density, cumulative probability of
    Mach numbers $> 1$, RMS Mach number, RMS vorticity, mass-weighted RMS vorticity, Taylor scale, 
    mass-weighted, Taylor scale, small-scale compressive ratio and the corresponding mass-weighted
    compressive ratio
    \label{tab:statistics}}
  \end{center}
\end{table*}

Regardless of the refinement variables in use, one has to find an appropriate normalisation in order to
set thresholds for triggering refinement. Whereas \citet{KritNor07} normalise the control variables in their AMR simulation with the local flow velocity and gas density, we determine thresholds for refinement from statistical properties of the associated control variables. This has the advantage that no \emph{a priori} knowledge about the control variables is required. For the calculation of the thresholds, both the spatial average $\langle h_{i}(\vect{x},t)\rangle$ of the control variable $h_{i}$ and its standard deviation $\mathrm{std}\,h_{i}(t)$, where $\mathrm{std}^{2}h_{i}(t)=\langle h_{i}^{2}(\vect{x},t)\rangle-\langle h_{i}(\vect{x},t)\rangle^{2}$, are used. A cell will be flagged for refinement, if the fluctuation
\begin{equation}
    \label{eq:refine}
    h_{i}^{\prime}=h_{i}-\langle h_{i}(\vect{x},t)\rangle\ge\mathrm{thresh}\,h_{i}(t),
\end{equation}
where the threshold is defined by 
\begin{equation}
  \mathrm{thresh}\,h_{i}(t) := A\max\left[
    \langle h_{i}(\vect{x},t)\rangle,
    \mathrm{std}\,h_{i}(t)\right].
\end{equation}
This definition of the threshold ensures that refinement is only triggered by pronounced fluctuations. For a nearly uniform field with $h_{i}^{\prime}(\vect{x},t)\ll \langle h_{i}(\vect{x},t)\rangle$, the above criterion amounts to $h_{i}^{\prime}(\vect{x},t)\gtrsim A\langle h_{i}(\vect{x},t)\rangle_{i}$,
i.~e., virtually no cells are refined provided that $A\gtrsim 1$. For the opposing limit of
a field with large fluctuations relative to the average, the refinement criterion becomes $h_{i}^{\prime}(\vect{x},t)\ge A\,\mathrm{std}\,h_{i}(t)$. In the general case of multiple levels of refinement, the averaging has to be constrained to refined regions or grid patches at higher levels. The efficiency of refinement can be tuned by the coefficient $A$. If $A\gg 1$, only peak fluctuation will trigger refinement. From test simulations with various choices of $A$, we concluded that statistical properties of turbulence are well reproduced for $A=1$, as one would intuitively expect. This setting that was used for the production run. 

\begin{figure*}[t]
  \begin{center}
    \mbox{\subfigure[$t=0.44T$, $n_{1}=4298$, $\chi_{1}=0.299$]{\includegraphics[width=8.5cm]{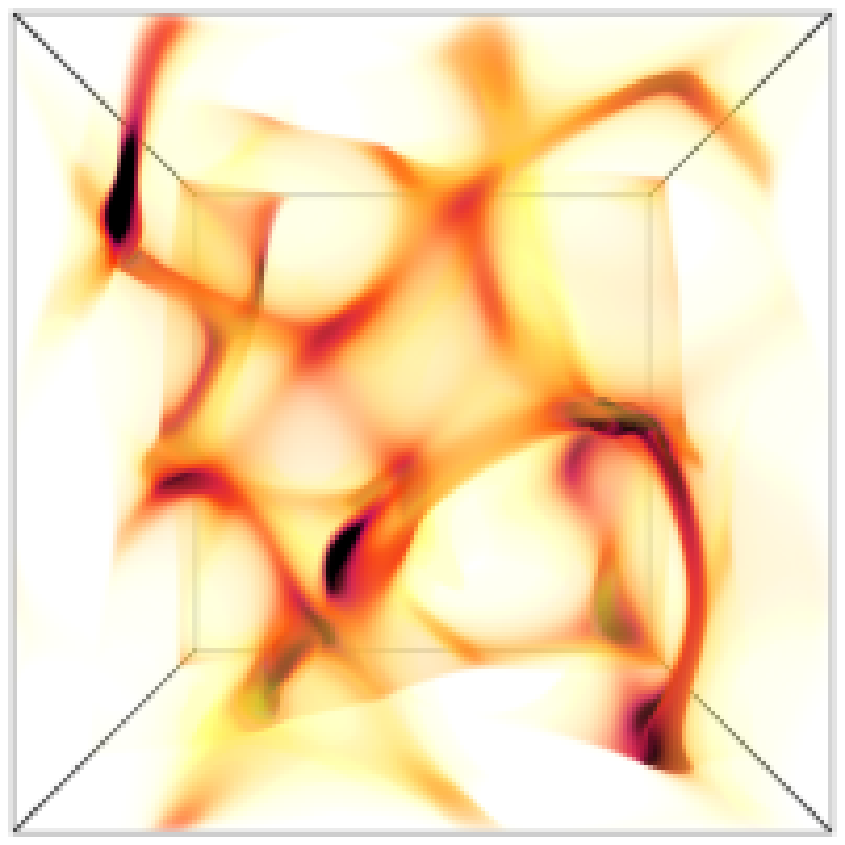}}\qquad
          \subfigure[$t=0.58T$, $n_{1}=5208$, $\chi_{1}=0.321$]{\includegraphics[width=8.5cm]{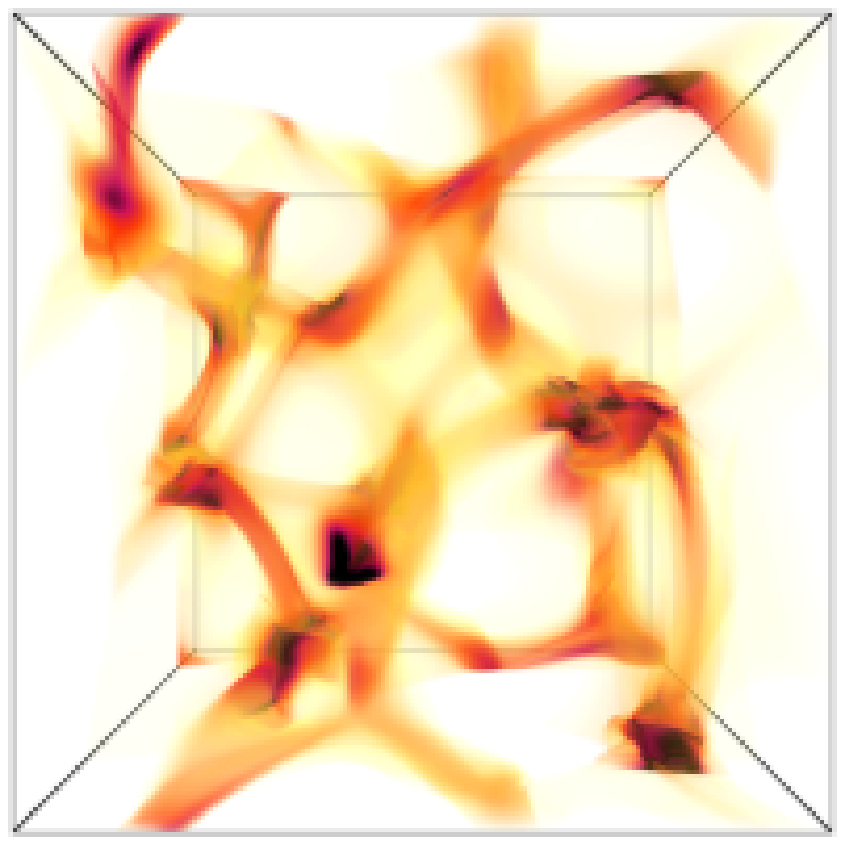}}}
    \mbox{\subfigure[$t=0.73T$, $n_{1}=6331$, $\chi_{1}=0.423$]{\includegraphics[width=8.5cm]{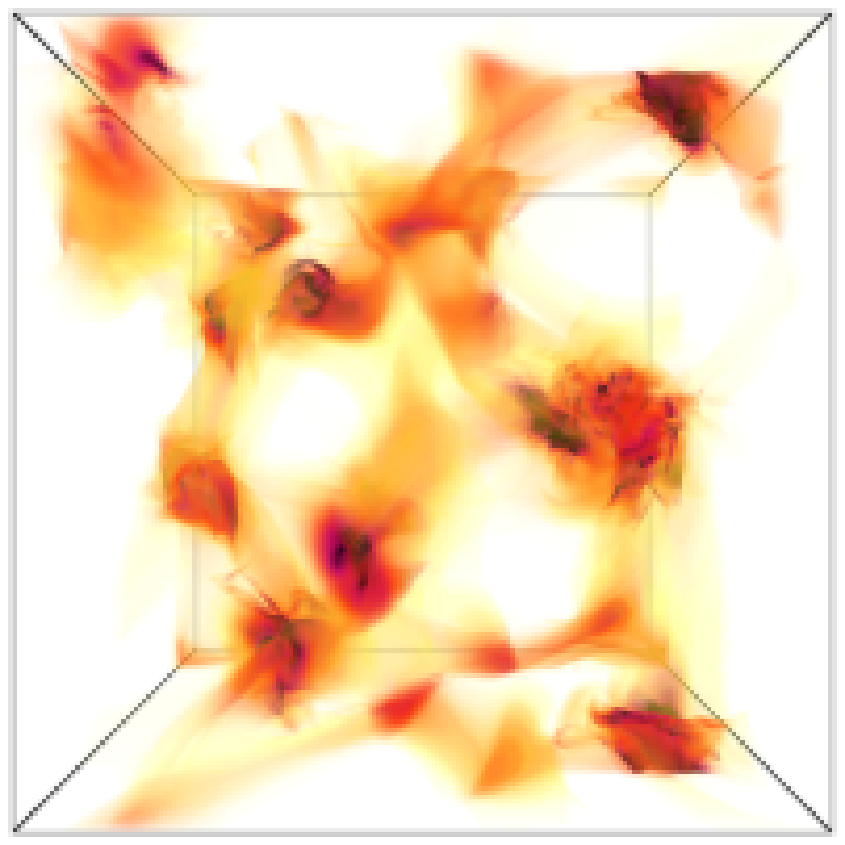}}\qquad
          \subfigure[$t=1.02T$, $n_{1}=8015$, $\chi_{1}=0.653$]{\includegraphics[width=8.5cm]{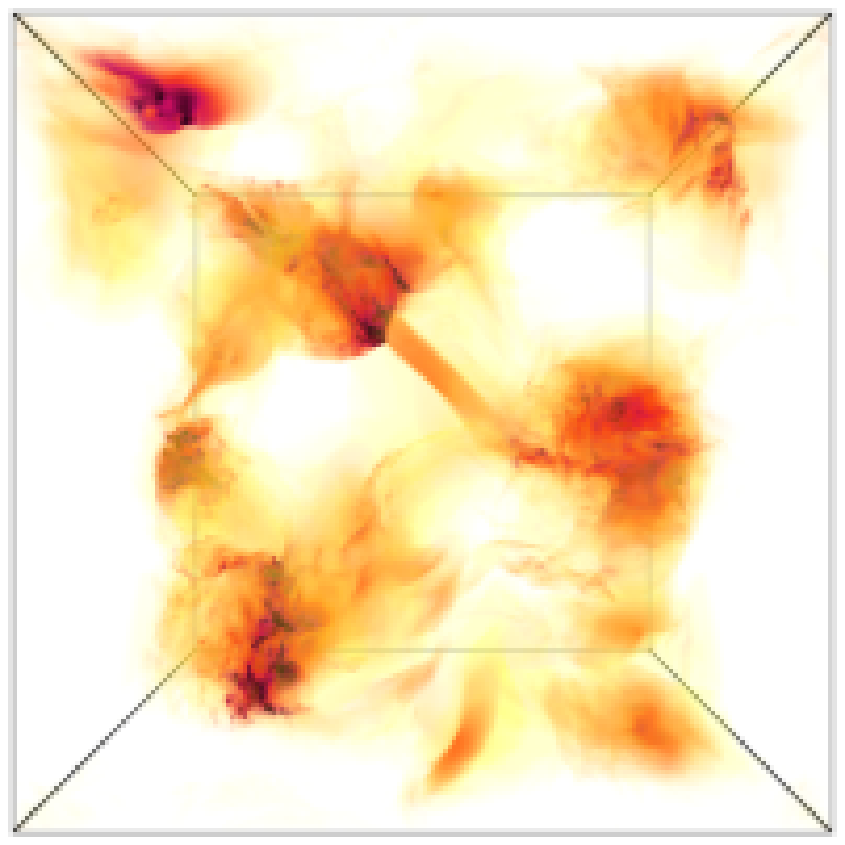}}}
    \caption{Volume renderings of the mass density $\rho$ for different stages of developing turbulence in the AMR simulation. For each instant of time, the total number of grid patches at the first level of refinement, $N_{1}$,  and the corresponding volume filling factor $\chi_{1}$ are specified. }
    \label{fg:dens3d}
  \end{center}
\end{figure*}

\begin{figure*}[t]
  \begin{center}
    \mbox{\subfigure[$t=0.44T$, $n_{1}=4298$, $\chi_{1}=0.299$]{\includegraphics[width=8.5cm]{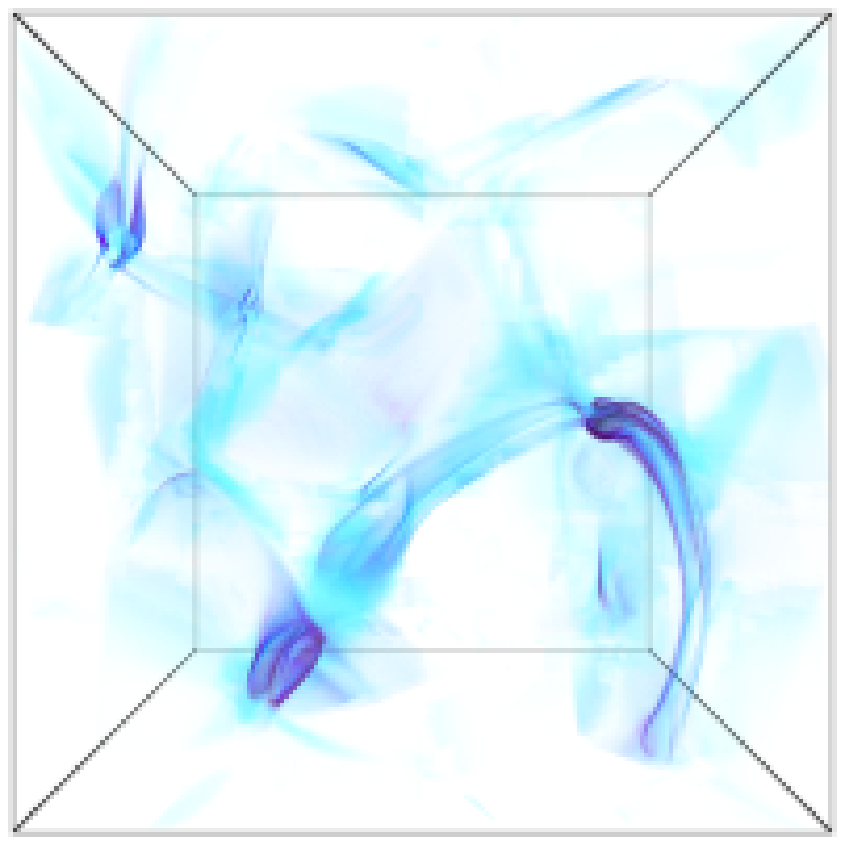}}\qquad
          \subfigure[$t=0.58T$, $n_{1}=5208$, $\chi_{1}=0.321$]{\includegraphics[width=8.5cm]{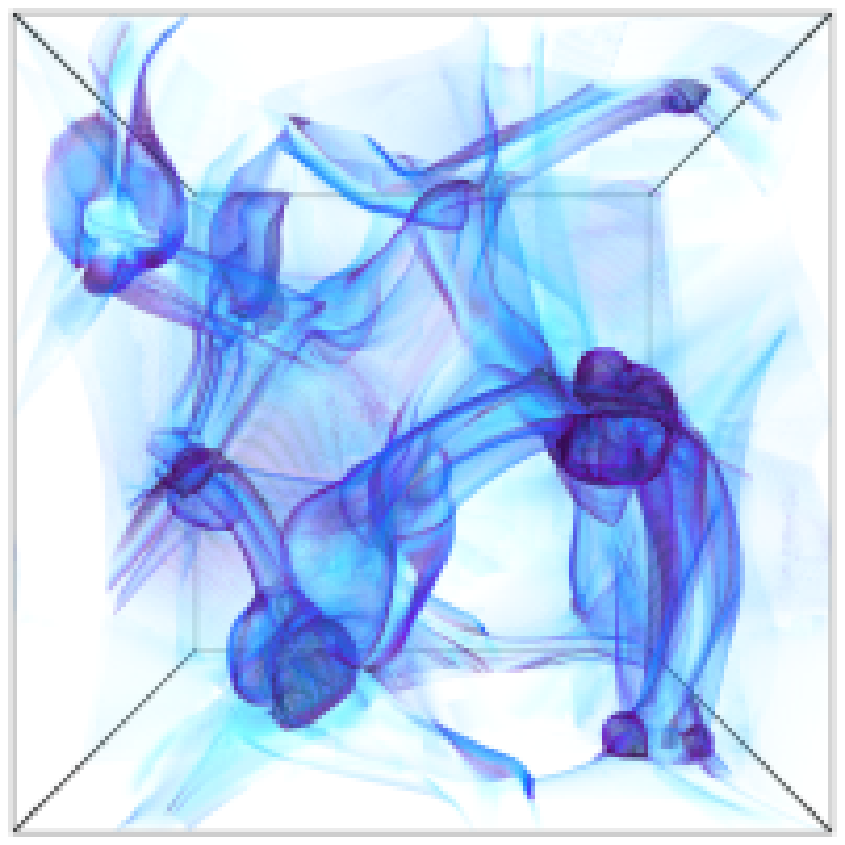}}}
    \mbox{\subfigure[$t=0.73T$, $n_{1}=6331$, $\chi_{1}=0.423$]{\includegraphics[width=8.5cm]{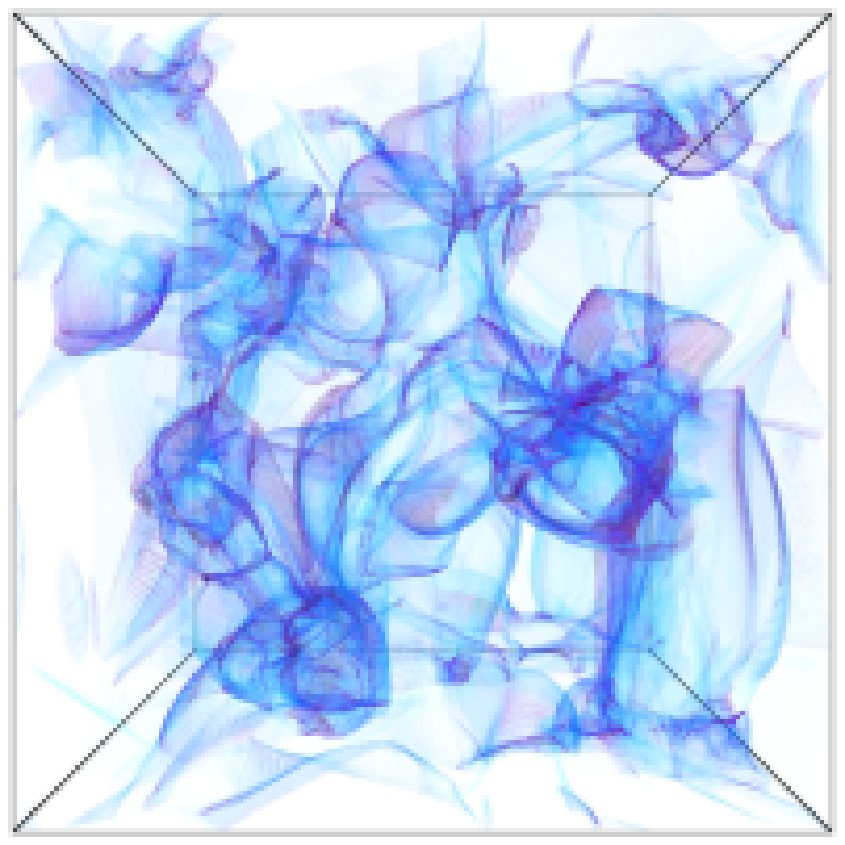}}\qquad
          \subfigure[$t=1.02T$, $n_{1}=8015$, $\chi_{1}=0.653$]{\includegraphics[width=8.5cm]{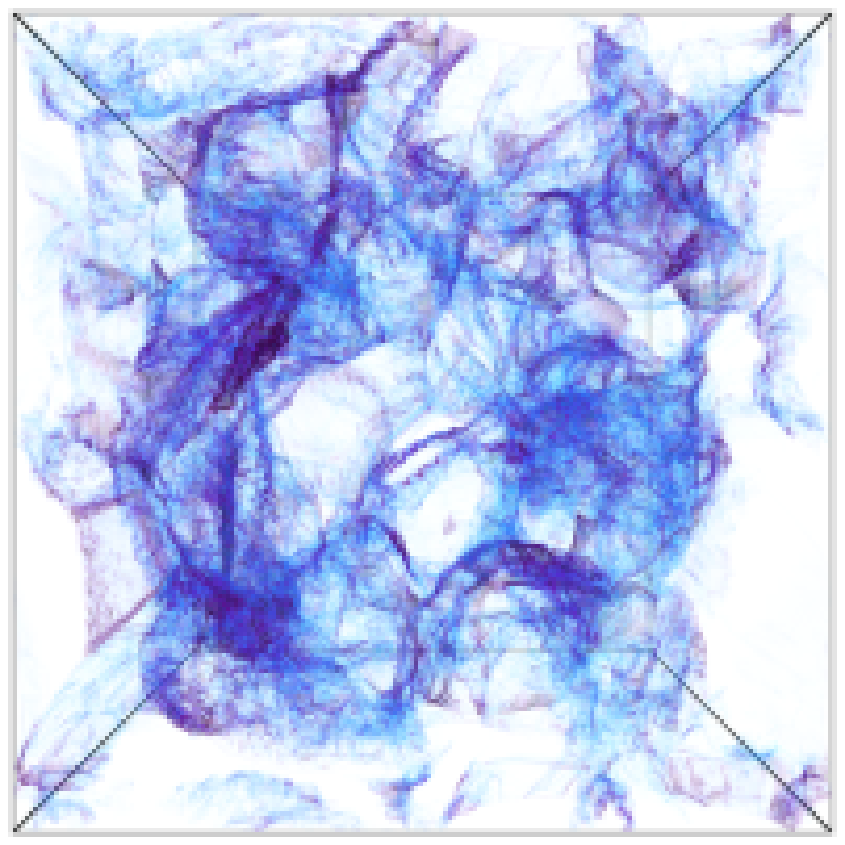}}}
    \caption{Volume renderings of the vorticity modulus $\omega$ corresponding to Figure~\ref{fg:dens3d}}
    \label{fg:vort3d}
  \end{center}
\end{figure*}

\begin{figure*}[t]
  \begin{center}
    \mbox{\subfigure[$\rho(t=0.44T)$]{\includegraphics[width=8.5cm]{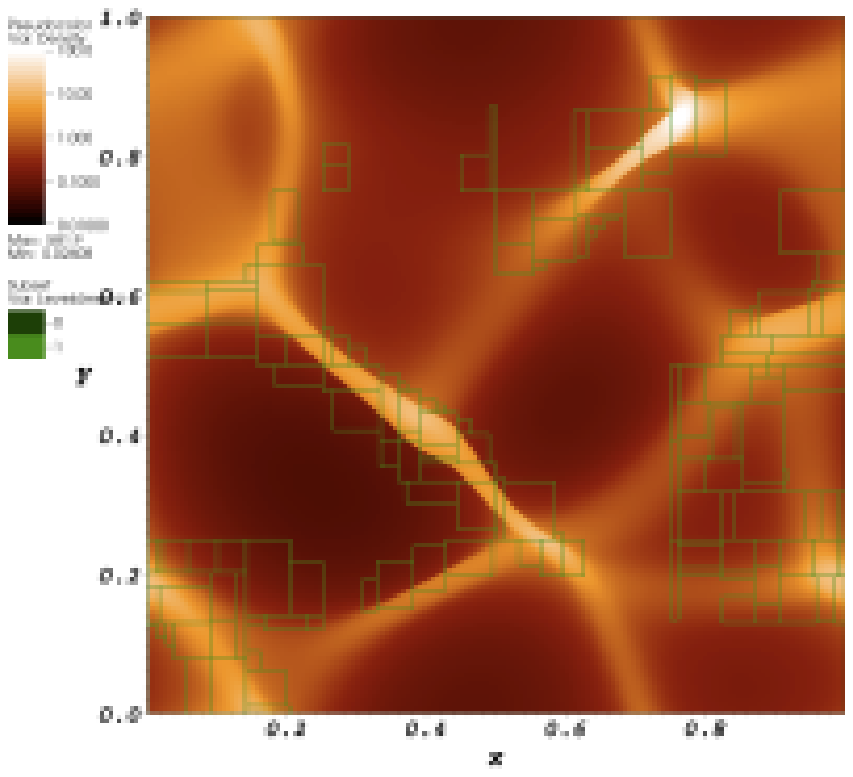}}\qquad
          \subfigure[$\mathcal{M}(t=0.44T)$]{\includegraphics[width=8.5cm]{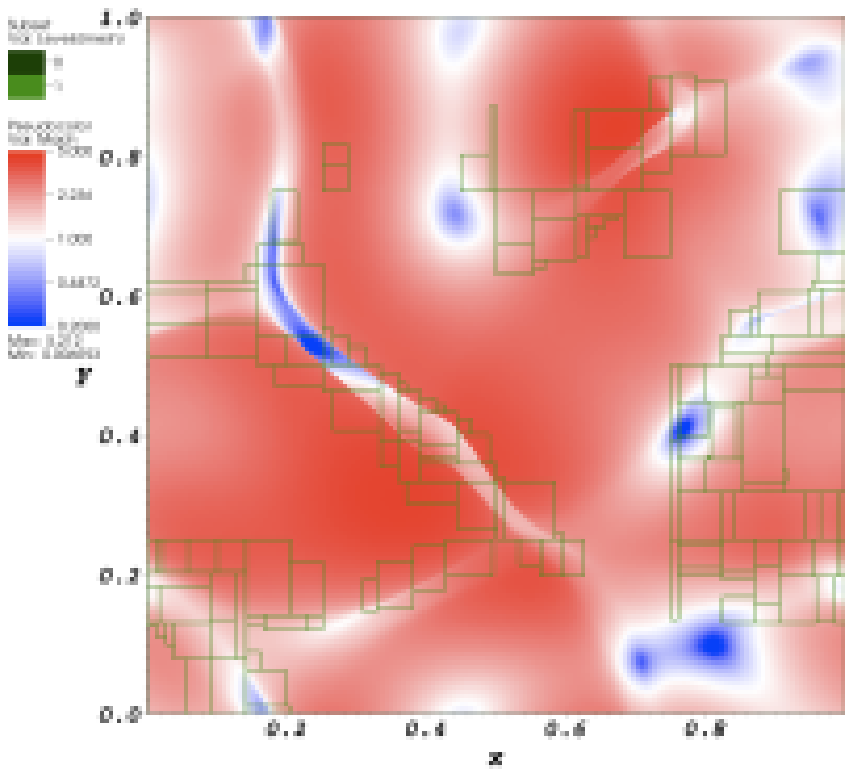}}}
    \mbox{\subfigure[$\rho(t=0.58T)$]{\includegraphics[width=8.5cm]{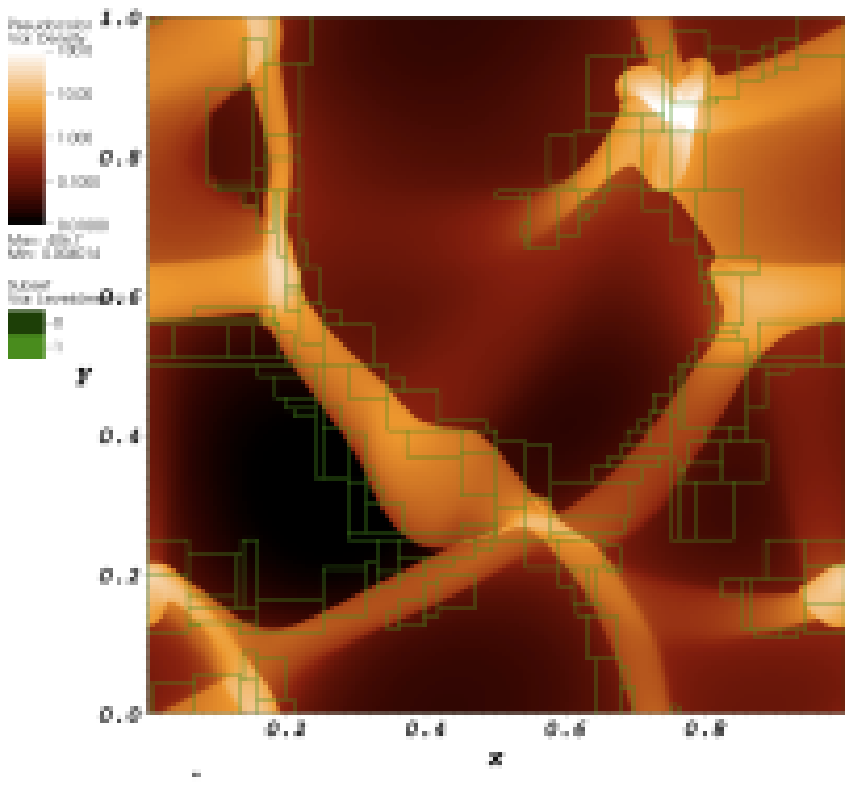}}\qquad
          \subfigure[$\mathcal{M}(t=0.58T)$]{\includegraphics[width=8.5cm]{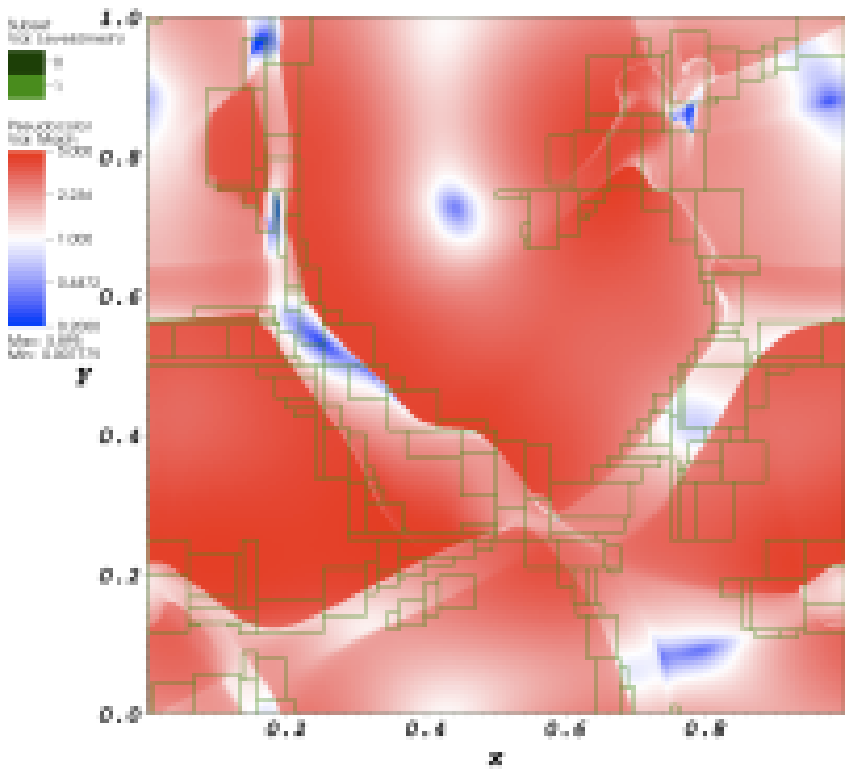}}}
    \caption{Contour plots of the mass density $\rho$ and the Mach number $\mathcal{M}$ in
    the plane $z=0$ corresponding to the top panels in Figures~\ref{fg:dens3d} and~\ref{fg:vort3d}.
    The rectangles show the boundaries of refined grid patches.}
    \label{fg:slice_a}
  \end{center}
\end{figure*}

\begin{figure*}[t]
  \begin{center}
    \mbox{\subfigure[$\rho(t=0.73T)$]{\includegraphics[width=8.5cm]{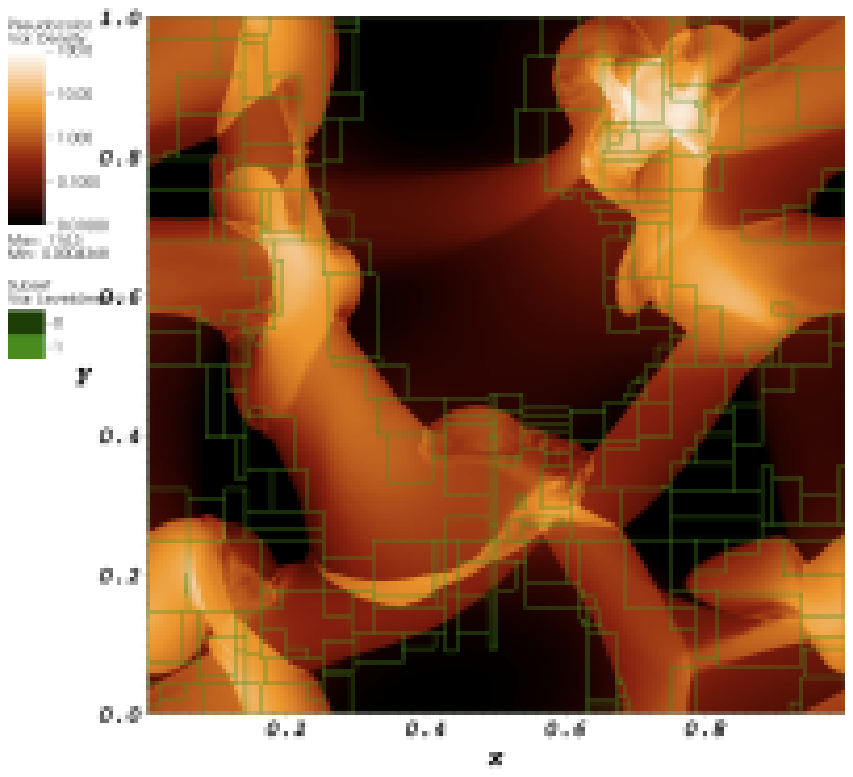}}\qquad
          \subfigure[$\mathcal{M}(t=0.73T)$]{\includegraphics[width=8.5cm]{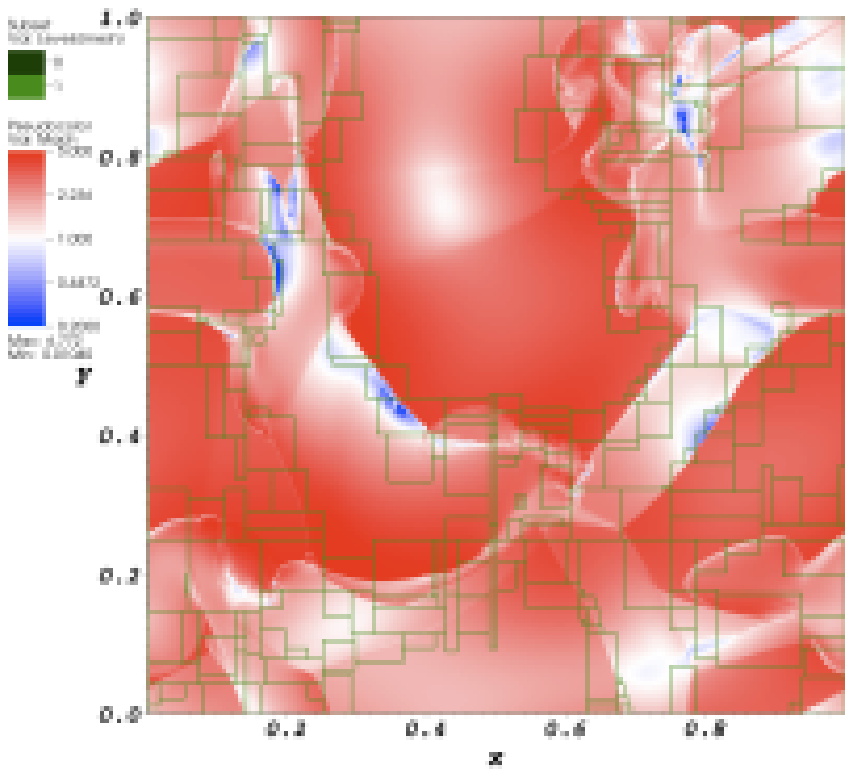}}}
    \mbox{\subfigure[$\rho(t=1.02T)$]{\includegraphics[width=8.5cm]{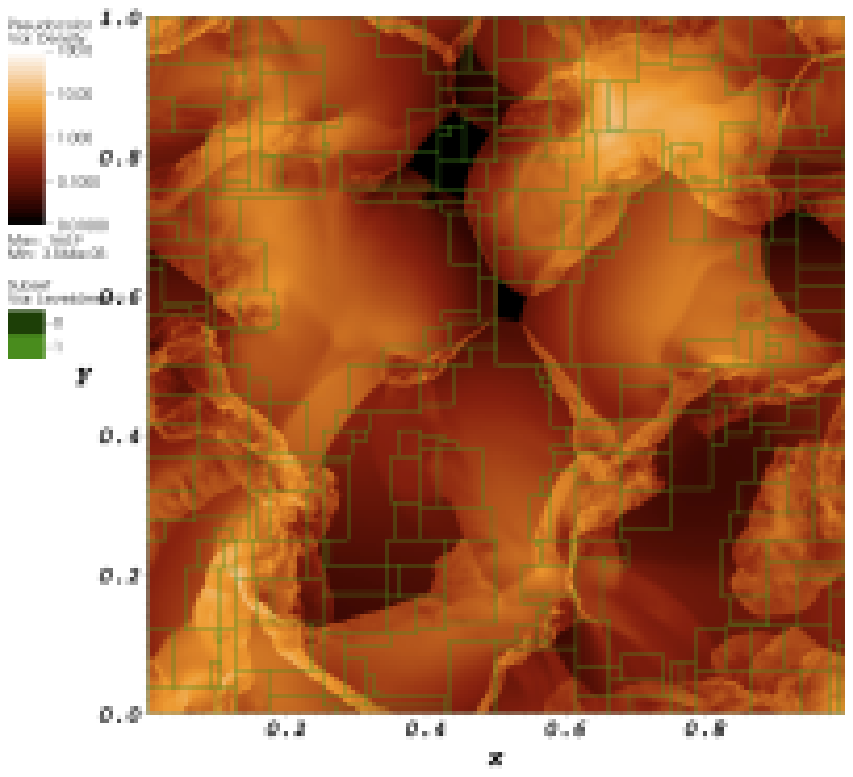}}\qquad
          \subfigure[$\mathcal{M}(t=1.02T)$]{\includegraphics[width=8.5cm]{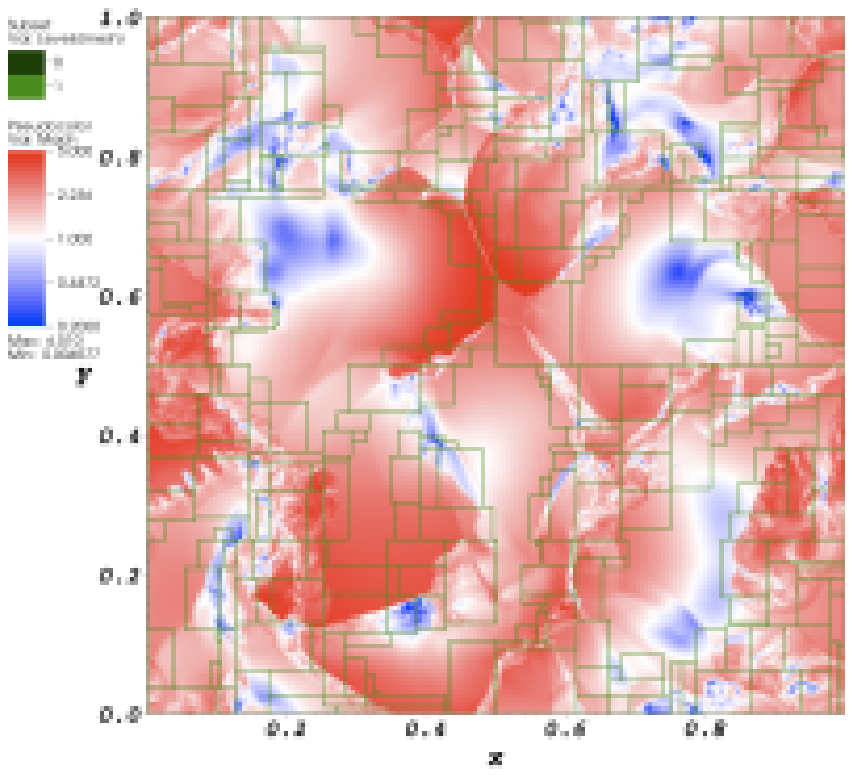}}}
    \caption{Contour plots of the mass density $\rho$ and the Mach number $\mathcal{M}$ in
    the plane $z=0$ corresponding to the bottom panels in Figures~\ref{fg:dens3d} and~\ref{fg:vort3d}. The rectangles show the boundaries of refined grid patches}
    \label{fg:slice_b}
  \end{center}
\end{figure*}

\section{Simulation results}
\label{sc:results}

Making use of the invariance properties of isothermal turbulence, we normalise all variables in the following such that the mean density $\rho_{0}=1$, the initial pressure $P_{0}=1$ and the size of the computational domain $X=2L=1$. This corresponds to the internal representation of the fluid-dynamical variables in Enzo. The speed of sound is then given by
$c_{0}=\sqrt{\gamma}\approx 1.005$, the characteristic velocity $V\approx 2.901$ and the integral time scale $T\approx 0.172$. The static-grid simulation run for $0\le t\le 1.6$, spanning about 9 integral times scales. Output was produced in intervals of $0.025$ resulting in 7 data dumps per integral time. An overview of global mean quantities is given in Table~\ref{tab:statistics}. For an illustration of the evolution within the first integral time scale, three-dimensional volume renderings of the mass density $\rho$ and the vorticity modulus $\omega$ at time $t/T=0.44$, $0.58$, $0.73$ and $1.02$ are shown in Figures~\ref{fg:dens3d} and~\ref{fg:vort3d}, respectively. These volume renderings were produced from AMR data. The AMR simulation was stopped when the first integral time scale was completed because, at this point, it progressed slower than the static-grid run. In part, this was caused by a severe performance drop of Enzo as the total number of grid patches approached $10000$.\footnote{Meanwhile, this problem has been resolved by the Enzo developer.} Apart from that, more than $60\,\%$ of the computational domain is covered by refined grid patches at time $t/T\approx 1$. With a single level of refinement, it is difficult to compete with a static-grid simulation for
such a volume filling factor. The development of the mesh structure in two-dimensional slices is illustrated in Figures~\ref{fg:slice_a} and~\ref{fg:slice_b}. Later in this Section, we will show that statistical properties inferred from both runs agree very well. In the conclusion of this article, we will comment on perspectives of applying AMR to the numerical simulation of turbulence. As regards the evolution over several integral time scales, we will refer to the static-grid data in the following.

Initially, as one can see from the maximal values and the standard deviations in Table~\ref{tab:statistics}, the mass density rises rapidly
and reaches a peak roughly at half an integral time scale. The RMS Mach number, $\mathcal{M}_{\mathrm{rms}}=\langle(v/c_{\mathrm{s}})^{2}\rangle^{1/2}$, is then close to $3$. The reason for the fast production of density enhancements of order $10^{3}$ is illustrated by panels (a) and (b) in Figure~\ref{fg:dens3d}.
The outflows of gas from proximate regions, in which the divergence of the stochastic force is positive, collide and become supersonic on a time scale $\sim T/\mathrm{Ma}\approx 0.2T$. Interpolation of the cumulative probabilities for Mach numbers greater than unity implies that about $60\,\%$ of the domain is filled by supersonic flows at time $t=0.2T$. The collision of these supersonic flows results in the sheetlike structures of shock-compressed gas that can be seen in the volume renderings. 
The density structure is compared to the Mach number $\mathcal{M}=v/c_{\mathrm{s}}$ of the gas in the two-dimensional slices plotted in Figures~\ref{fg:slice_a} and~\ref{fg:slice_b}. Since the forcing evolves on a slower time scale than the colliding flows, gas accumulates in the over-dense regions faster than the drift of the positive-divergence regions could act against further compression. On the other hand, compressed gas re-expands in between reflected shock fronts travelling in opposite directions. Presently, the gas becomes turbulent due to instabilities at the collision interfaces and the solenoidal forcing component. As one can see in Figures~\ref{fg:slice_a} and~\ref{fg:slice_b}, these interfaces are well tracked by AMR, while smooth regions of over-dense gas are not generally refined. At the end of the first integral time, cloud-like structures have emanated from the vertices of the high-density sheets (see panels (c) and (d) in Figures~\ref{fg:dens3d}). The maximal densities are somewhat smaller at this stage and $\mathcal{M}_{\mathrm{rms}}\approx 2.5$. Subsequently, $\mathcal{M}_{\mathrm{rms}}$ slowly decreases because of the gradual increase of the internal energy caused by energy dissipation.

A crucial question for the following analysis is whether turbulence settles into a steady state, in which small-scale scale velocity fluctuations are fully developed and kinetic energy dissipation is balanced by the integral-scale supply of energy from the forcing. To this end, we consider several statistical measures. The Taylor scale $\lambda$ follows from the two-point autocorrelation of the turbulent velocity field. For incompressible isotropic turbulence, $\lambda$ is given by the
the time scale associated with the RMS vorticity, $1/\omega_{\mathrm{rms}}$, times the RMS velocity fluctuation \citep{Frisch}:
\begin{equation}
	\label{eq:taylor}
	\lambda=\sqrt{5}\frac{u_{\mathrm{rms}}}{\omega_{\mathrm{rms}}}.
\end{equation}
Using the above expression as definition of $\lambda$ also in the case of compressible turbulence,
we calculated the values listed in Table~\ref{tab:statistics}. Initially, $\lambda\sim L$, indicating that the flow is smooth and only varies over the integral length $L$. In the course of the first integral time, the Taylor scale drops substantially and then gradually approaches the asymptotic value $\lambda\approx 20$. The early evolution of $\lambda$ reflects the growth of small-scale structure that can be seen in Figure~\ref{fg:vort3d}. As velocity fluctuations develop at smaller scales, the vorticity rises rapidly and, hence, $\lambda$ decreases. 

Based on the equilibrium between energy injection and dissipation in the statistically stationary state, the Taylor scale can be related to the Reynolds number \citep{Pope}:
\begin{equation}
	\frac{\lambda}{L}=\sqrt{\frac{10}{\mathrm{Re}}}.
\end{equation}
This relation offers the possibility of estimating the Reynolds number from statistical properties of the velocity field without explicit knowledge of the numerical viscosity. Substituting $L=384\Delta$ and $\lambda/\Delta\approx 20$, we obtain $\mathrm{Re}\approx 3.7\cdot 10^{3}$. Since the criterion $\mathrm{Re}=(L/\Delta)^{4/3}$ yields $\mathrm{Re}\approx 2.8\cdot 10^{3}$, 
the Reynolds number in our simulation is comparable to the Reynolds number that could be achieved in a simulation of incompressible turbulence with the same dynamical range.

Since the Taylor scale only probes the vorticity of the velocity field, we also consider the so-called small-scale compressive ratio \citep{KiOrs90},
\begin{equation}
	r_{\mathrm{cs}}=\frac{d_{\mathrm{rms}}^{2}}{d_{\mathrm{rms}}^{2}+\omega_{\mathrm{rms}}^{2}}.
\end{equation}
At early time, this ratio is close to unity, i.~e., the rotation-free component of the velocity dominates. This 
reflects the large-scale stochastic force. Subsequently, however, vorticity develops in the vicinity of shock-fronts and grows due to the production of turbulence. The compressive ratio in the steady state is slightly less than $0.5$. Thus, the solenoidal and dilatational modes of the velocity fluctuations are comparable in magnitude. For comparison,
\citet{KritNor07} found $r_{\mathrm{cs}} \approx 0.28$ in their simulation.

The definitions of the Taylor scale~(\ref{eq:taylor}) and the small-scale compressive ratio~(\ref{eq:compr_ratio}) stem from the theory of incompressible turbulence. We also calculated mass-weighted variants based
on the kinetic energy density, $\frac{1}{2}\rho u^{2}$, the density of the enstrophy, $\frac{1}{2}\rho\omega^{2}$, and $\frac{1}{2}\rho d^{2}$. The values of the mass-weighted Taylor scale,
\begin{equation}
	\lambda_{\mathrm{mw}}=\sqrt{5\frac{\langle\rho u^{2}\rangle}{\langle\rho\omega^{2}\rangle}},
\end{equation}
listed in Table~\ref{tab:statistics} show a faster decline in the course of the first integral time in comparison to $\lambda$. This suggests that a significant fraction of the vorticity originates from shock-compressed gas. Indeed, panels (a) to (c) in Figure~\ref{fg:vort3d} show sheet-like structures
which can be attributed to shock fronts at the interface between compressed and rarefied gas.
Moreover, Figure \ref{fg:slice_b} suggests that the production of turbulence ensues in the vicinity of shocks. For $t/T\ge 2$, we have $\lambda_{\mathrm{mw}}/\lambda\approx 0.6$. Note that the intermittency of the mass density causes stronger fluctuations of $\lambda_{\mathrm{mw}}$. The mass-weighted small-scale compressive ratio,
\begin{equation}
	\label{eq:compr_ratio}
	r_{\mathrm{mwcs}}= \frac{\langle\rho d^{2}\rangle}{\langle\rho\omega^{2}+\rho d^{2}\rangle},
\end{equation}
asymptotically approaches a value slightly less than $0.3\approx 0.6 r_{\mathrm{cs}}$, indicating that
solenoidal flow is somewhat more pronounced relative to dilatational flow in over-dense gas. Altogether,
the evolution of the Taylor scale and the small-scale compressive ratio shows that it takes about
two time scales for the turbulent flow to become statistically stationary, possibly, with small systematic changes up to $t/T\approx 4$.

In the remainder of this Section, the statistics of the mass density and the turbulent velocity field will be
analysed further by means of probability density functions, spectrum functions and structure functions.

\begin{figure}[t]
  \begin{center}
    \resizebox{\hsize}{!}{\includegraphics[angle=90]{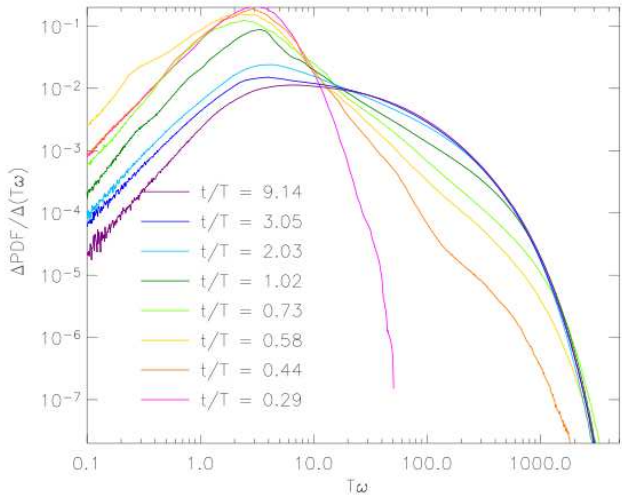}}
    \caption{Temporal evolution of the probability density functions of the vorticity modulus $\omega$.}
    \label{fg:vort_pdfs}
  \end{center}
\end{figure}

\begin{figure}[t]
  \begin{center}
     \resizebox{\hsize}{!}{\includegraphics[angle=90]{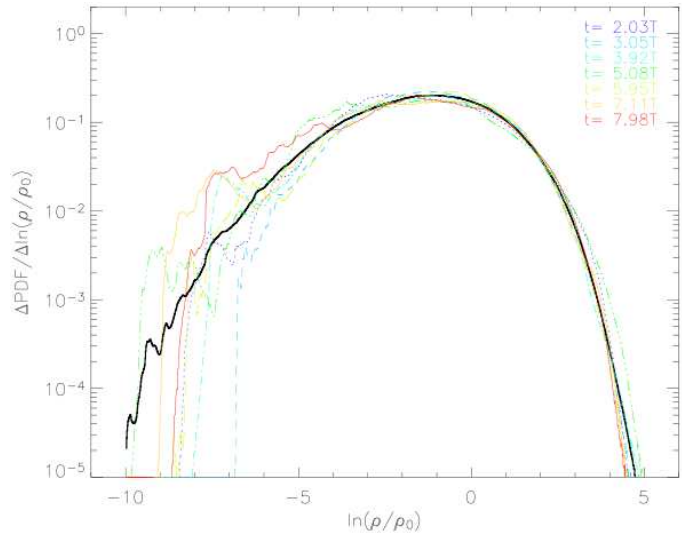}}
    \caption{Instantaneous and temporally averaged (thick solid line) logarithmic mass density fluctuations, $\mathrm{ln}(\rho/\rho_{0})$.}
    \label{fg:dens_pdf}
  \end{center}
\end{figure}

\subsection{Probability density functions}
\label{sc:pdfs}

For statistical analysis, turbulent flow quantities such as the gas density
can be considered as spatiotemporal random variables. In numerical
simulations based on finite-volume schemes, Eulerian statistics is most readily
obtained by calculating probability distributions for the whole computational
domain at any given instant of time. The probability distribution function (PDF)
quantifies the cumulative probability that a random variable is found to be less than a
certain value. The derivative of the probability distribution function with respect to value of the
random variable is called the probability density function (pdf).\footnote{Here, we adhere to the convention that the symbol of the derivative is written in lower case letters, while the primitive function is written in upper case letters.} In the following, velocity and mass density statistics is specified by means of probability density functions.

The pdfs of the vorticity modulus plotted in panel (b) of
Figure~\ref{fg:vort_pdfs} highlight the rapid evolution of
the flow within the first integral time. At $t\approx 0.5T$, 
a power-law decrease of the $\omega$-pdf in the subrange
$10\lesssim\omega\lesssim 100$ becomes apparent and
flattens subsequently. The power-law behaviour points at a self-similar
ensemble of shocks. The flattening arises form the formation
of small-scale structure, as illustrated in Figure~\ref{fg:vort3d},
because velocity fluctuations at smaller scales are associated with higher vorticity.
At later time, the power-law subrange
diminishes and the exponential tail towards high vorticity, which
is characteristic for incompressible turbulence, becomes more prominent. 
This indicates a transition from an early shock-dominated phase into
vortex-dominated turbulence in the course of the first few integral times. 
From $t\approx 3T$ onwards, the $\omega$-pdfs show only little variation in time.

The logarithmic density fluctuations, $\delta=\ln(\rho/\rho_{0})$, on the other hand,
show no systematic evolution in the aftermath of the shock-dominated
production phase. This corresponds to the nearly constant RMS Mach number
for $t/T>1$ (see Table~\ref{tab:statistics}). In Figure~\ref{fg:dens_pdf},
$\delta$-pdfs are plotted for several instants of time together
with a time-averaged pdf. For the averaging $57$ data sets produced in the time interval $1 \le t/T\le 9$ were used. One can see random variations of the instantaneous pdfs, which
reflect intermittent deviations from the ensemble average. The average of the
pdfs over several integral times, on the other hand, yields a sound approximation to
the statistics of the mass density in the ensemble average. Most noticeably, the averaged $\delta$-pdf is not symmetric but exhibits negative skewness, i.~e., lower densities are more probable in relation to higher densities. This corresponds to the appearance of compact regions of over-dense gas surrounded by extended voids in Figure\ref{fg:dens3d}.

\begin{figure}[t]
  \begin{center}
    \resizebox{\hsize}{!}{\includegraphics{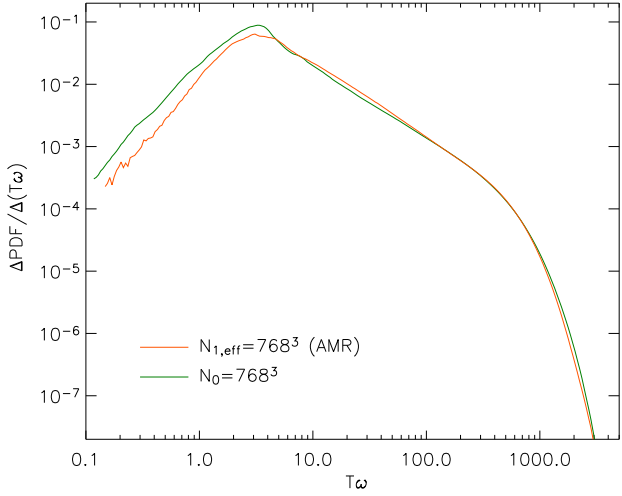}}
    \caption{Comparison of the probability density functions of the vorticity $\omega$ for the static grid and the AMR run at time $t=1.02T$.}
    \label{fg:vort_pdf_amr}
  \end{center}
\end{figure}

\begin{figure}[t]
  \begin{center}
    \resizebox{\hsize}{!}{\includegraphics{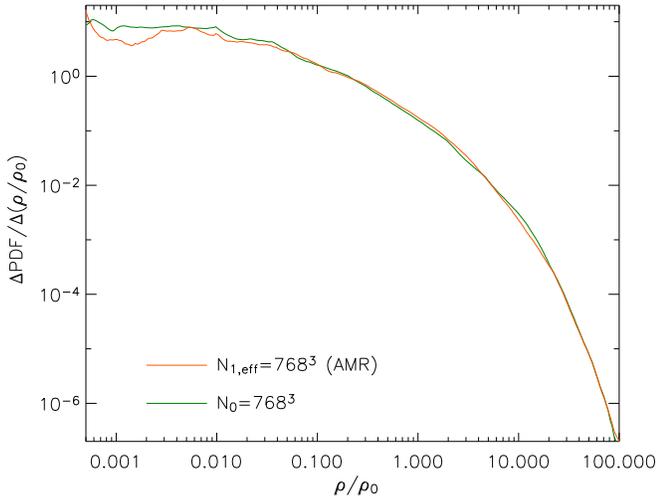}}
    \caption{Comparison of the probability density functions of the mass density $\rho$ for the static grid and the AMR run at time $t=1.02T$.}
    \label{fg:dens_pdf_amr}
  \end{center}
\end{figure}

\begin{figure}[t]
  \begin{center}
    \resizebox{\hsize}{!}{\includegraphics{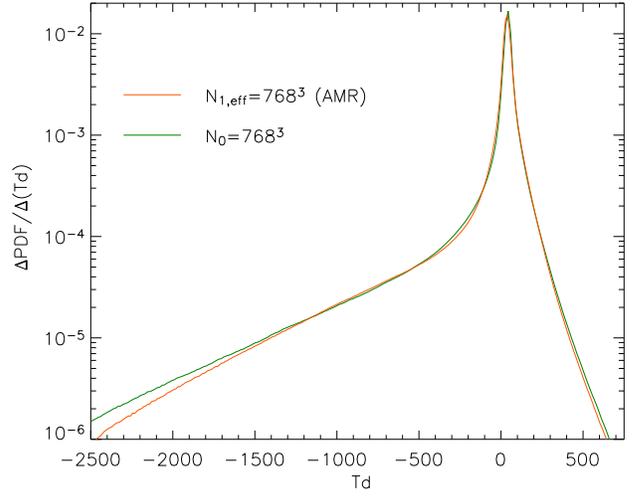}}
    \caption{Comparison of the probability density functions of the divergence $d$ for the static grid and the AMR run at time $t=1.02T$.}
    \label{fg:div_pdf_amr}
  \end{center}
\end{figure}

Probability density functions are indicators for the reliability of AMR simulations,
because pdfs would be affected, if the fluctuations of the corresponding quantities
were not properly tracked by refinement.  Here, we consider pdfs of the mass density, the vorticity modulus and divergence. For the termination time of the AMR simulation ($t=1.02T$),
Figures~\ref{fg:dens_pdf_amr}, \ref{fg:vort_pdf_amr} and \ref{fg:div_pdf_amr} show 
that the pdfs calculated from the AMR data are well matched by the corresponding static-grid
pdfs.\footnote{Note that the pdf of $\rho$ is related to the pdf of $\delta$ via
$\rho\,\mathrm{pdf}(\rho)=\mathrm{pdf}(\delta)$. } In particular, the high-density and
high-vorticity tails arising, respectively, from strongly compressed
gas and from thin vortex filaments are very well reproduced with AMR. There are
deviations in the left tails of the pdfs though. This is to be expected because the
refinement criteria are optimised to capture peak values that contribute to the
tails on the right. The important physics is associated with these tails. Since turbulence is close
to statistical equilibrium after the first integral time scale, the comparison demonstrates that, in principle, the AMR techniques used in our simulation are applicable to turbulence simulations.
For conclusive testing, however, an improved code version is necessary to carry out
AMR simulations with multiple levels of refinement over several integral time scales. 

\subsection{Turbulence energy spectra} \label{sc:spectra}

The turbulence energy spectrum function \citep{Frisch},
\begin{equation}
	E(k)=\frac{1}{2}\oint\dd\Omega_{k}\,k^{2}\hat{\vect{u}}(k)\cdot\hat{\vect{u}}^{\ast}(k)
\end{equation}
where $\hat{\vect{u}}(k)$ is the Fourier transform of the velocity field and $\hat{\vect{u}}^{\ast}(k)$
its complex conjugate, is the surface integral of the spectral energy density per unit mass over
a sphere of radius $k$ in Fourier space. In the ensemble average, the energy spectrum function of statistically stationary turbulence is independent of time and obeys a power law,
$E(k) \propto k^{-\beta}$, for $k_{0}=2\pi/L\ll k\ll k_{\mathrm{c}}$, where $k_{\mathrm{c}}=\pi/\Delta$ is the numerical cutoff wave number.  We computed discrete energy
spectrum functions form several data sets following the prescription of \citet{SchmHille06}. 
The resulting spectrum functions were compensated with the factor $k^{2}$ so that the compensated spectrum of Burgers turbulence appears constant in the inertial subrange. Moreover,
the normalization $k=1$ for the smallest wave number (corresponding to the box size) is applied.

\begin{figure*}[t]
  \begin{center}
    \includegraphics[width=17cm]{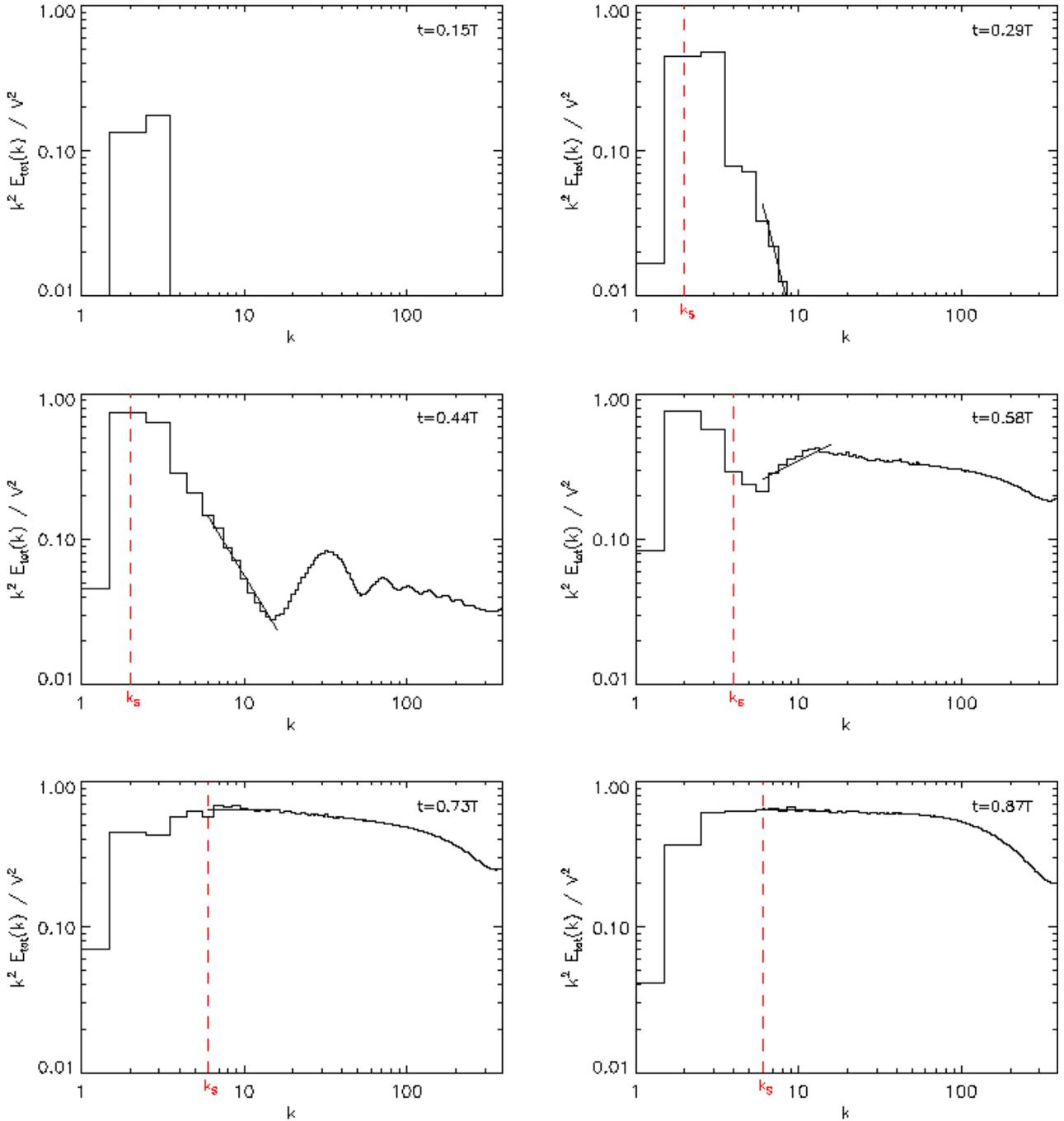}
    \caption{Early evolution of the energy spectra for  $t<T$. The spectrum functions
    are compensated with $k^{2}$. The sonic wave number is
    indicated by vertical dotted lines. Additionally, the Kolmogorov slope -5/3 is depicted as
    dashed line.}
    \label{pic:energyspectra1}
  \end{center}
\end{figure*}

\begin{figure*}[t]
  \begin{center}
    \includegraphics[width=17cm]{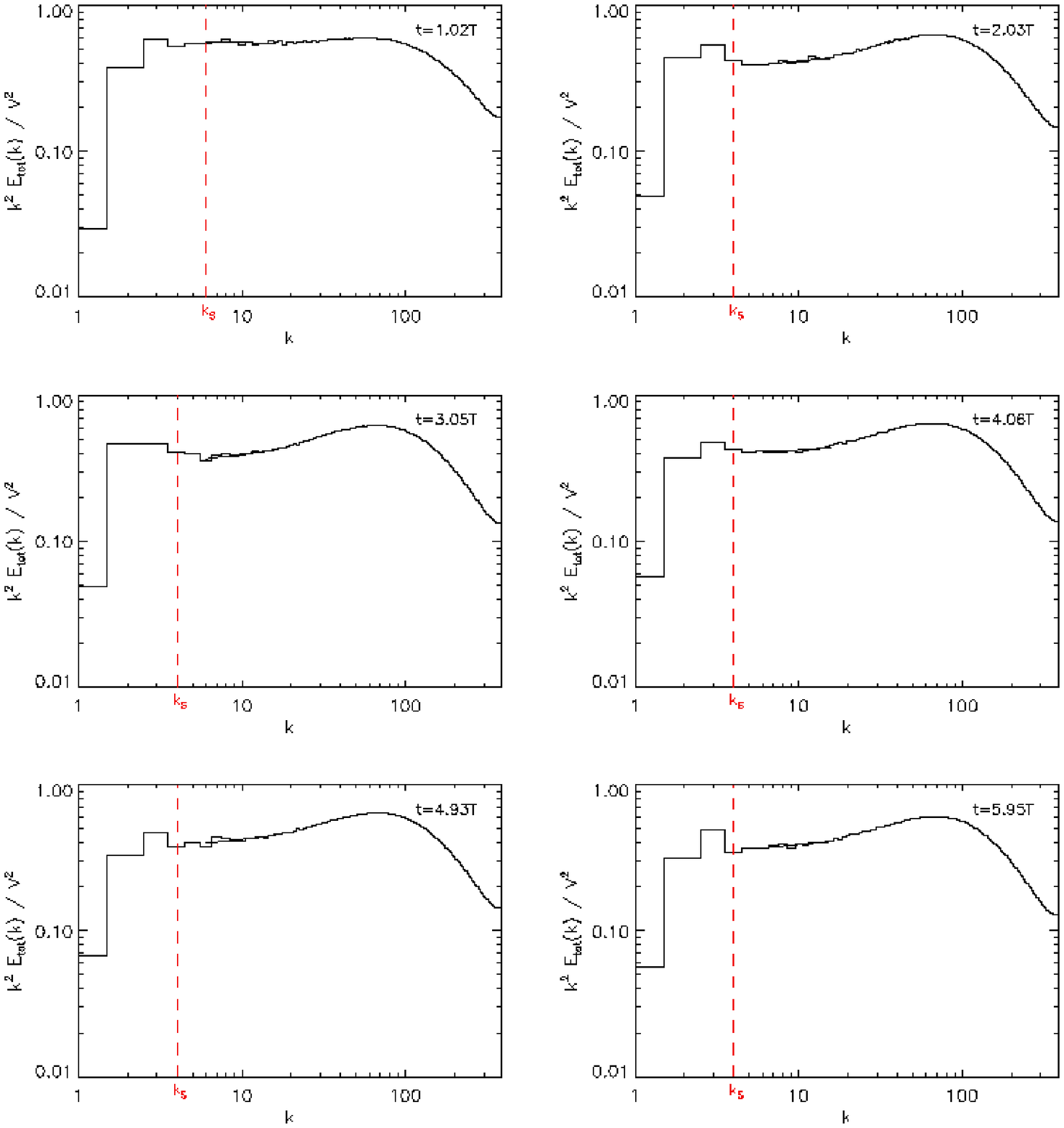}
    \caption{Temporal evolution of the compensated energy spectra continuing
    Figure~\ref{pic:energyspectra1} for $t>T$.}
    \label{pic:energyspectra2}
  \end{center}
\end{figure*}

The sequence of compensated energy spectrum functions for consecutive instants
of time shown in Figures~\ref{pic:energyspectra1}
and~\ref{pic:energyspectra2}, illustrates the gradual build-up of
the turbulent energy cascade in the course of the first integral
time scale followed by the relaxation of the flow into a steady state. In the following,
it is understood that the spectral indices $\beta$ are obtained by fitting power laws $k^{\beta'}$ to the compensated spectrum functions, where $\beta'=\beta-2$.
The spectra for $t\approx T$ are substantially steeper than the Kolmogorov spectrum $E(k) \propto k^{-5/3}$. In particular, we find $\beta\approx-2.02 \pm 0.06$ at time $t=1.02\,T$, which is
very close to $\beta=2$ for Burgers turbulence and verifies an
early shock-dominated phase. Subsequently, the spectrum functions become flatter and 
a bump roughly centred at $k\approx 70\approx k_{\mathrm{c}}/5$ appears. This bump, which stems from the so-called bottleneck
effect \citep{Dobler+03,SchmHille06}, severely constraints the power-law range of the spectra.
Applying, least square fits in the wave number range $5 < k < 15$  yields
the time-averaged estimate $\beta\approx 1.87$ for $t/T\ge 2$ (see Table~\ref{tab:indices}).  However, due to the gradual flattening toward higher wave numbers caused by the bottleneck effect and, possibly, other factors related to PPM \citep{KritNor07}, 
the systematic uncertainty of this estimate is much higher than the fitting error margins.


The integral of $E(k)$ over the range of numerically resolved wavenumbers is the mean 
kinetic energy per unit mass:
\begin{equation}
  \label{eq:parseval_energy}
  \int_{k_{0}/\alpha}^{k_{\mathrm{c}}} E(k)\dd k =
  \frac{1}{2}u_{\mathrm{rms}}^2,
\end{equation}
where $\alpha=X/L$ and $k_{0}=2\pi/L$ (see Section~\ref{sc:numerics}).
For supersonic isothermal turbulence, $u_{\mathrm{rms}}>c_{0}$.
Hence, there exists a wave number $k_{\mathrm{s}}>k_{0}/\alpha$, for which
the cumulative turbulence energy in the range $k\gtrsim
k_{\mathrm{s}}$ equals $\frac{1}{2}\langle
c_{\mathrm{s}}^2\rangle\simeq c_{0}^{2}$, i.~e.,
\begin{equation}
  \label{eq:sonic_wave}
  \int_{k_{\mathrm{s}}}^{k_{\mathrm{c}}} E(k)\dd k \simeq
  \frac{1}{2}\langle c_{\mathrm{s}}\rangle^2.
\end{equation}
The sonic wave number $k_{\mathrm{s}}$ corresponds to the length scale $\ell_{\mathrm{s}}=2\pi/k_{\mathrm{s}}$, for which the turbulent velocity fluctuations $u'(\ell_{\mathrm{s}})\sim c_{\mathrm{s}}$.
The values of $k_{\mathrm{s}}$ obtained  by the inversion of equation~(\ref{eq:sonic_wave})
are  marked by vertical lines in the plots of the energy spectrum functions 
(Figures~\ref{pic:energyspectra1} and \ref{pic:energyspectra2}). 
In the course of the first integral time, $k_{\mathrm{s}}$ reaches a maximum of about $3k_{0}$.
Together with the maximum of $k_{\mathrm{s}}$ comes the steepest slope with a spectral index of
$\beta\approx 2.0$. Afterwards, the sonic wave number settles at $2k_{0}$. One should note that
no implications can be made with regard to the scaling for $k>k_{\mathrm{s}}$. Although it would
seem that this subrange of wave numbers is dominated by nearly incompressible modes, the scaling
is probably not consistent with the Kolmogorov theory. Since the uncertainty in the determination of 
$\beta$ is quite high due to the bottleneck effect, a stronger case for this conclusion will be made on grounds of the exponent of the second-order structure functions.

\begin{table}[h]
  \begin{center}
      \begin{tabular}{c|cc}\hline
$t/T$ & $\beta$ & $\beta_{\delta}$ \\ \hline\hline
0.73 & $1.99 \pm 0.10$ &                               \\ \hline
0.87 & $2.04 \pm 0.05$ &                               \\ \hline
1.02 & $2.02 \pm 0.06$ &                               \\ \hline
2.03 & $1.89 \pm 0.05$ & $1.85 \pm 0.06$ \\ \hline
3.05 & $1.85 \pm 0.05$ & $1.76 \pm 0.04$ \\ \hline
4.06 & $1.92 \pm 0.04$ & $1.75 \pm 0.05$ \\ \hline
4.93 & $1.87 \pm 0.08$ & $1.82 \pm 0.05$ \\ \hline
5.95 & $1.90 \pm 0.04$ & $1.80 \pm 0.04$ \\ \hline
      \end{tabular}
  \end{center}
  \caption{Spectral indices of turbulence energy and mass density fluctuation spectra with
  	fit error margins.
           \label{tab:indices}}
\end{table}

\begin{figure}[t]
  \begin{center}
    \resizebox{\hsize}{!}{\includegraphics{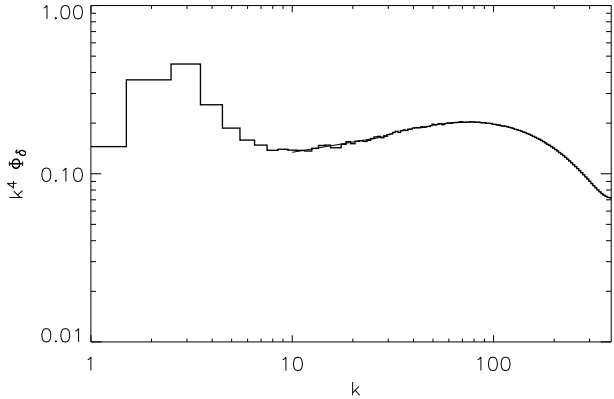}}
    \caption{Compensated spectral density of the quadratic logarithmic density fluctuation at time $t/T=5.95$.}
    \label{fg:delta_spectr}
  \end{center}
\end{figure}

\subsection{Mass density spectra}

Analogous to $\Phi(\vect{k})=\hat{\vect{u}}(k)\cdot\hat{\vect{u}}(k)^{\ast}$,
which specifies the density of specific kinetic energy, $\frac{1}{2}u^{2}$, in Fourier space, the spectral density
of the squared logarithmic density fluctuation $\delta^{2}$ is given by
\begin{equation}
	\Phi_{\delta}(\vect{k}) =\hat{\delta}(k)\cdot\hat{\delta}^{\ast}(k),
\end{equation}
and, by virtue of Parseval's theorem,
\begin{equation}
	\int\dd^{3}k\,\Phi_{\delta}(\vect{k}) = \langle\delta^{2}\rangle=\delta_{\mathrm{rms}}^{2}.
\end{equation}
For statistically isotropic turbulence, the spectral density $\Phi_{f}(\vect{k})$ of some field variable $f$ is a function of the wave number $k=|\vect{k}|$ only. The spectral density $\Phi_{\delta}(k)$ is the counterpart to $\Phi(k)$ in as much as both are related to squared fluctuations of the respective field variable. The numerically computed spectral density $\Phi_{\delta}(k)$ compensated with $k^{4}$ is plotted in Figure~\ref{fg:delta_spectr}. Similar to the turbulence energy spectra, there is only a marginal inertial subrange. Form least-square fits of the compensated spectral density $\Phi_{\delta}(k)$ in the range $10 < k < 30$, we obtained the spectral indices $\beta_{\delta}$ of the quadratic logarithmic density fluctuations. Averaging the indices listed in Table~\ref{tab:indices} for $t/T\approx 2$ to $6$ yields $\beta_{\delta}\approx 1.80$.

\subsection{Velocity structure functions}

The turbulence energy spectrum function $E(k)$ specifies the scaling behaviour
of the squared velocity fluctuation in Fourier space. The
corresponding description in physical space is given by the second-order
structure function, i.~e., the spatial two-point correlation function of the
velocity field. Structure functions of order $p$ are defined by
\begin{equation}
	S_{p}(\ell) =
	\langle|\vect{u}(\vect{x}+\vect{\ell}) - \vect{u}(\vect{x})|^p\rangle 
	\sim \ell^{\,\zeta_{p}},
\label{eq:structurefunctions}
\end{equation}
where $\ell=2\pi/k$ specifies the length scale. Of particular importance are the exponents 
$\zeta_{2}$ and $\zeta_{3}$ which are related to the scaling of turbulence energy and the turbulence energy flux, respectively. \citet{KritNor07} proposed to calculate structure functions for
the mass-weighted velocity $\rho^{1/3}\vect{u}$. This is motivated by the
dependence of the energy flux on $\rho u^{3}$ in compressible turbulence. However, observational
results for the scaling of molecular cloud turbulence are only available from measurements of the velocity dispersion at different scales. For this reason, we restrict our discussion to the
scaling properties of the structure functions without mass-weighing~(\ref{eq:structurefunctions}).

\begin{figure*}[t]
  \begin{center}
    \mbox{\subfigure[]{\includegraphics[width=8.5cm]{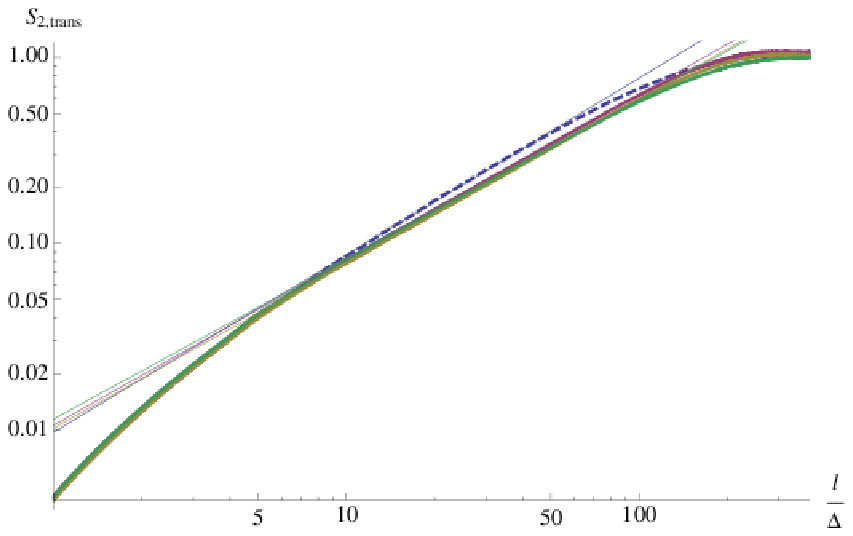}}\qquad
          \subfigure[]{\includegraphics[width=8.5cm]{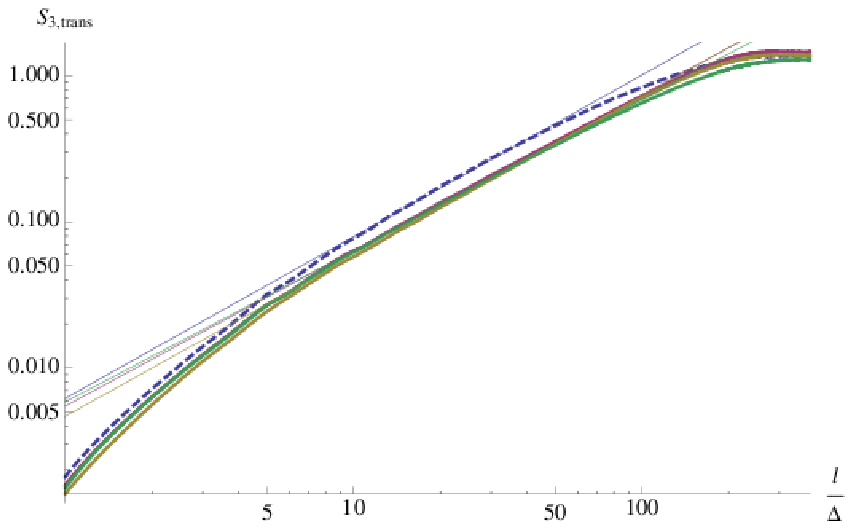}}}
    \caption{Second-order (a) and third-order (b) transversal structure functions
    	for $t/T\approx 1$ (dashed lines) and $t/T\approx 2,3,4$ (solid lines), respectively.
	Also shown are the corresponding power-law fit functions (thin lines). The structure functions are normalised by $V^{p}$.}
    \label{pic:structurefunctions}
  \end{center}
\end{figure*}

The computation of statistically converged structure functions of higher order is a
computationally demanding task. We applied a Monte-Carlo algorithm to sum
velocity differences from randomly sampled points over bins of spatial
separation.  All results presented hereafter have been computed based on approximately $2.4
\times 10^{10}$ data pairs. To ensure that sufficient numbers of
data pairs were sampled, we increased the amount of sampled data by
a factor of $5$ in one particular case. No significant differences were found for the structure functions up to fifth order. Since the measurement of line broadening at positions separated by a certain angular distance within molecular clouds corresponds to transversal
velocity fluctuations, we computed transversal structure functions, $S_{p,\mathrm{trans}}(\ell)$, for which the velocity increments $\vect{u}(\vect{x}+\vect{\ell}) - \vect{u}$ are projected perpendicular to $\vect{\ell}$. For $t/T\approx 1$, $2$, $3$ and $4$, the structure functions of second order and third order are plotted in Figure~\ref{pic:structurefunctions}. We fitted power laws in the subrange $8\leq \ell/\Delta \leq 50$ for $t/T\approx 1$ and $8 \leq \ell/\Delta \leq 70$  for $t/T\ge 2$, respectively, which are also shown in Figure~\ref{pic:structurefunctions}. Table~\ref{tab:powerexponents} lists the resulting scaling exponents $\zeta_{p}$ for $p=1,\ldots,5$ in intervals
of about one integral time scale. As one can see, $S_{2,\mathrm{trans}}$ is steeper at time $t/T\approx 1$ in comparison to later instants of time. The scaling exponent $\zeta_{2}\approx 0.95$ is close to the exponent of Burgers turbulence. For $t/T > 2$, there appears to be little variation of $S_{2,\mathrm{trans}}$, although there might be a slight systematic decrease of the scaling exponents till $t/T\approx 4$. 

\begin{table*}[t]
  \begin{center}
      \begin{tabular}{c|ccccc}\hline
$t/T$ & $p=1$ & $p=2$ & $p=3$ & $p=4$ & $p=5$\\ \hline\hline
1.02 & 0.654 $\pm$ 0.002 & 0.950 $\pm$ 0.003 & 1.107 $\pm$ 0.006 &
1.202 $\pm$ 0.009 & 1.256 $\pm$ 0.012\\ \hline
2.03 & 0.562 $\pm$ 0.004 & 0.888 $\pm$ 0.003 & 1.066 $\pm$ 0.008 &
1.167 $\pm$ 0.013 & 1.228 $\pm$ 0.016\\ \hline
3.05 & 0.538 $\pm$ 0.005 & 0.884 $\pm$ 0.003 & 1.097 $\pm$ 0.006 &
1.239 $\pm$ 0.013 & 1.344 $\pm$ 0.019\\ \hline
4.06 & 0.531 $\pm$ 0.003 & 0.858 $\pm$ 0.004 & 1.035 $\pm$ 0.013 &
1.123 $\pm$ 0.023 & 1.163 $\pm$ 0.031\\ \hline
4.93 & 0.530 $\pm$ 0.002 & 0.848 $\pm$ 0.006 & 1.018 $\pm$ 0.016 &
1.100 $\pm$ 0.026 & 1.135 $\pm$ 0.034\\ \hline
5.95 & 0.523 $\pm$ 0.003 & 0.852 $\pm$ 0.004 & 1.044 $\pm$ 0.012 &
1.157 $\pm$ 0.021 & 1.224 $\pm$ 0.027\\ \hline
6.96 & 0.546 $\pm$ 0.004 & 0.875 $\pm$ 0.004 & 1.053 $\pm$ 0.010 &
1.150 $\pm$ 0.017 & 1.208 $\pm$ 0.021\\ \hline
      \end{tabular}
  \end{center}
  \caption{Scaling exponents $\zeta_{p}$ obtained from power-law fits
  	to the structure functions $S_{p,\mathrm{trans}}(\ell)$ for
           several stages of the turbulent flow evolution.
           \label{tab:powerexponents}}
\end{table*}

Plotting $S_{2,\mathrm{trans}}$ against $S_{3,\mathrm{trans}}$, as shown in panel (a) of Figure~\ref{pic:ESS}, reveals power-law behaviour that partially extends into the dissipation range and also applies to the largest scales $\ell\sim L$. This remarkable property is 
known as extended self-similarity (ESS) \citep{BenzCil93}. For $t/T\approx 2$, the structure functions $S_{p,\mathrm{trans}}$ are shown as functions of $S_{3,\mathrm{trans}}$ for $p=1,\ldots,5$ in panel (b) of Figure~\ref{pic:ESS}. It appears that ESS is very well satisfied for the transversal structure functions with $p\le 5$, albeit with a somewhat smaller range in the case $p=5$. The slopes of the corresponding power-law fits functions for $0.04\le S_{3,\mathrm{trans}}\le 1.0$ yield relative scaling exponents $Z_{p}=\zeta_{p}/\zeta_{3}$. Averaging the values of $Z_{p}$ for $t/T\ge 2$, we obtained the values that are summarised in Table~\ref{tab:resultcomparison}. The averaged scaling
exponents are plotted with the corresponding scattering range for each $p$ in Figure~\ref{pic:scaling}.

\section{Discussion}
\label{sc:discussion}

\subsection{Velocity statistics}

The second order exponent $\zeta_{2}$ determines the inertial-range scaling of the RMS velocity fluctuations, $u'(\ell)\propto \ell^{\,\gamma}$, where $\gamma=\zeta_{2}/2$. From the power-law fits to transversal structure functions of second order for $t/T\ge 2$, we obtain the average $\zeta_{2}\approx 0.868$ corresponding to $\gamma\approx 0.43$, while $\gamma\approx 0.48$ is implied for $t/T\approx 1$. The spectral index $\beta=2\gamma+1$  (see Section~\ref{sc:spectra}) implies $\gamma\approx 0.5$ in the transient phase prior to equilibrium and $\gamma\approx 0.45$ for statistically stationary turbulence in our simulation. Given the large uncertainty
of fitting the turbulence energy spectra, the spectral indices are consistent with the scaling exponents obtained from the second-order structure functions. Observations of molecular clouds indicate power law relations between the velocity dispersion and the size, $\Delta v\propto \ell^{\,\alpha}$. The index $\alpha$ is interpreted to correspond more or less to $\gamma$, although there is not a strict equivalence \citep{BruntHey02b}. While \citet{FalPu92} found $\alpha\approx 0.4$ and $\alpha\approx 0.43$ was obtained by \citet{MieBal94} in very good agreement with the steady-state value of $\gamma$ inferred from our simulation, \citet{BruntHey02b} deduced from their measurements a much higher mean $\gamma\approx 0.57$. However, their results for individual molecular clouds vary in the range $\gamma\approx 0.33\ldots 0.81$. For Perseus, \citet{PadJuv06} obtained an index $\beta\approx 1.81$ corresponding to $\gamma\approx 0.4$, again very close to the index we estimated for the turbulence energy spectra in statistical equilibrium.

\begin{figure*}[t]
  \begin{center}
    \mbox{\subfigure[]{\includegraphics[width=8.5cm]{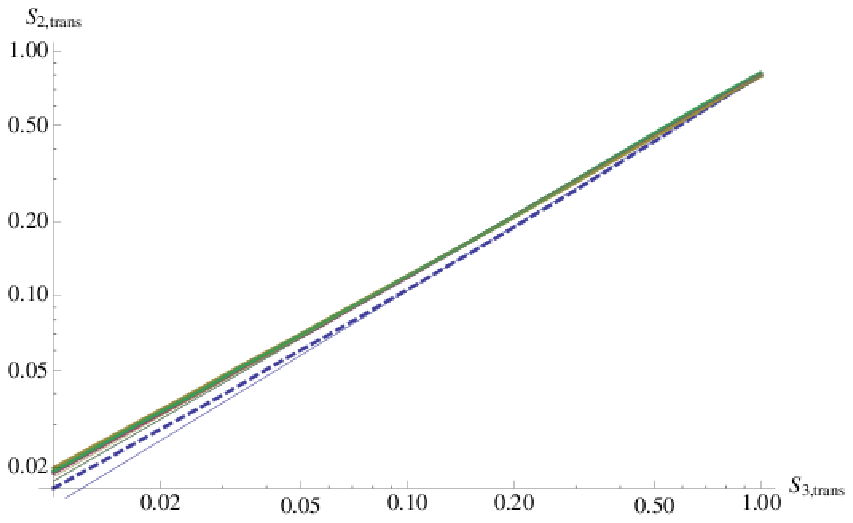}}\qquad
          \subfigure[]{\includegraphics[width=8.5cm]{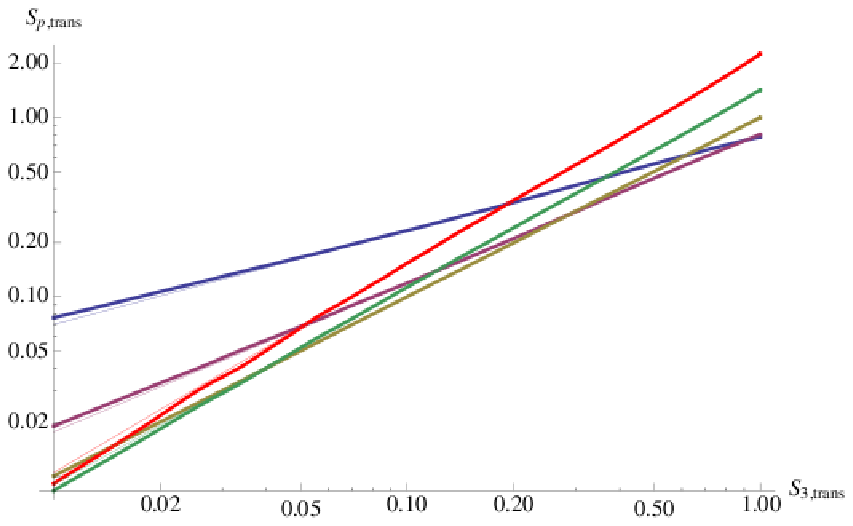}}}
    \caption{Transversal structure functions plotted against the corresponding third-order 
    structure functions. Panel (a) shows $S_{2,\mathrm{trans}}$ as
    function of $S_{3,\mathrm{trans}}$ for the same instants of time
    as in Figure~\ref{pic:structurefunctions}. In panel (b), structure
    functions (thick lines) and the corresponding power-law fits (thin lines) up to fifth order at time $t\approx 2.03T$ are shown. }
  \label{pic:ESS}
  \end{center}
\end{figure*}

Other than the numerical determination of spectral indices and absolute scaling exponents, relative scaling properties can be inferred quite reliably (see Figure~\ref{pic:ESS}). In Table~\ref{tab:resultcomparison}, various theoretical predictions and numerical results for the relative scaling exponents $Z_{p}=\zeta_{p}/\zeta_{3}$ of order $p=1,\ldots,5$ are listed. The classical Kolmogorov theory of statistically stationary isotropic incompressible turbulence yields $Z_{p}=\zeta_{p}=p/3$ \citep{Kolmog41,Frisch}. \citet{SheLeveque94} included intermittency corrections for incompressible turbulence based on the idea of a hierarchy of dissipative structures. The scaling exponents predicted by their model were experimentally confirmed with high precision. \citet{Dub94} showed that this model is a member of the general class of hierarchical structure models based on log-Poisson statistics. Consider the local rate of dissipation $\epsilon_{\ell}$ at some length scale $\ell$. Going to smaller length scales, the rate of dissipation tends to grow as structure at smaller scales emerges. This implies the scaling law $\epsilon_{\ell}\propto \ell^{-\Delta}$, where $\Delta$ is the scaling exponent of the \emph{most singular} dissipative structures associated with the divergent rate of dissipation as $\ell\rightarrow\infty$. On the other hand, intermittency tends to decrease the rate of dissipation. Let $\beta$ be a parameter (not to be confused with the spectral index of turbulence energy)  that specifies the degree of intermittency in the fashion of the $\beta$ model \citep{Frisch} and let us assume that the dissipation rates at different length scales are connected by $n$ events out of a Poisson distribution, which break up the dissipative structures from larger toward smaller length scales. Then, for any two length scales $\ell_{2}<\ell_{1}$,
\begin{equation}
	\epsilon_{\ell_{2}}=\left(\frac{\ell_{2}}{\ell_{1}}\right)^{-\Delta}\beta^{n}\epsilon_{\ell_{1}}.
\end{equation}
Because of the law of finite energy dissipation \citep{Frisch}, we have the normalisation constraint
$\langle\epsilon_{\ell_{2}}\rangle=\langle\epsilon_{\ell_{2}}\rangle=\epsilon$, where $\epsilon$ is the mean rate of energy dissipation. Invoking the Poisson distribution, it follows that
$\langle\epsilon_{\ell}^{p}\rangle\propto\ell^{-\Delta p+\Delta(1-\beta^{p})/(1-\beta)}$. With the refined similarity hypothesis \citep{Frisch}, the relative scalings 
\begin{equation}
	\label{eq:Z2par}
	Z_{p} =(1-\Delta)\frac{p}{3}+\frac{\Delta}{1-\beta}\left(1 - \beta^{p/3}\right) 
\end{equation}
for the velocity structure functions are obtained \citep{Dub94}.\footnote{Our simplified motivation follows \citet{PanWheel08}.} To determine $\Delta$, \citet{SheLeveque94} postulated a a uniform time scale $\propto\ell^{\Delta}$ for the dissipation of various turbulence intensities. The scaling of the most singular dissipative structures is given by the largest available kinetic energy, which is of the order $V^{2}$, divided by $\ell^{\Delta}$. In the case of incompressible turbulence, $\Delta=2/3$. Setting
$\beta=2/3$, they interpreted $C=\Delta/(1-\beta)=2$ as codimension of one-dimensional vortex filaments, which are considered to be the most singular dissipative structures of incompressible turbulence. 

\begin{table*}[t]
  \begin{center}
%    \begin{tiny}
      \begin{tabular}{c|ccc|ccc|cc}\hline
$p$ & K41 & SL94 & B02 & BNP02 & KNPW07 & this work & D94 & D94\\
 &  & ($C=2$) & ($C=1$) &  (MHD $500^{3}$) & (HD $1024^{3}$) & (HD $768^{3}$) & (C=0.71) &
 (C=1.23) \\ \hline\hline
1 & 0.33 & 0.36 & 0.42 & 0.42 & 0.43 & 0.52 & 0.54 & 0.53\\ \hline
2 & 0.67 & 0.70 & 0.74 & 0.74 & 0.76 & 0.83 & 0.82 & 0.83\\ \hline
3 & 1.00 & 1.00 & 1.00 & 1.00 & 1.00 & 1.00 & 1.00 & 1.00\\ \hline
4 & 1.33 & 1.28 & 1.21 & 1.20 &      & 1.09 & 1.14 & 1.10\\ \hline
5 & 1.67 & 1.54 & 1.40 & 1.38 &      & 1.14 & 1.26 & 1.16\\ \hline
      \end{tabular}
%    \end{tiny}
  \end{center}
  \caption{Comparison of the normalised scaling exponents $Z_{p}=\zeta_{p}/\zeta_{3}$ for structure
               functions of order $p=1,\ldots,5$ predicted by \citet{Kolmog41}, 
               \citet{SheLeveque94}, \citet{Boldyrev02} and the general hierarchical structure model
               of \cite{Dub94} with numerical results for transversal structure functions from 
               \citet{Boldyrev+02}, \citet{KritNor07} and this work. Note that $\mathcal{M}_\mathrm{rms}\approx 10$ for BNP02 and $\mathcal{M}_\mathrm{rms}\approx 6$ for KNPW07.
               \label{tab:resultcomparison}}
\end{table*}

\begin{figure}[t]
  \begin{center}
    \resizebox{\hsize}{!}{\includegraphics{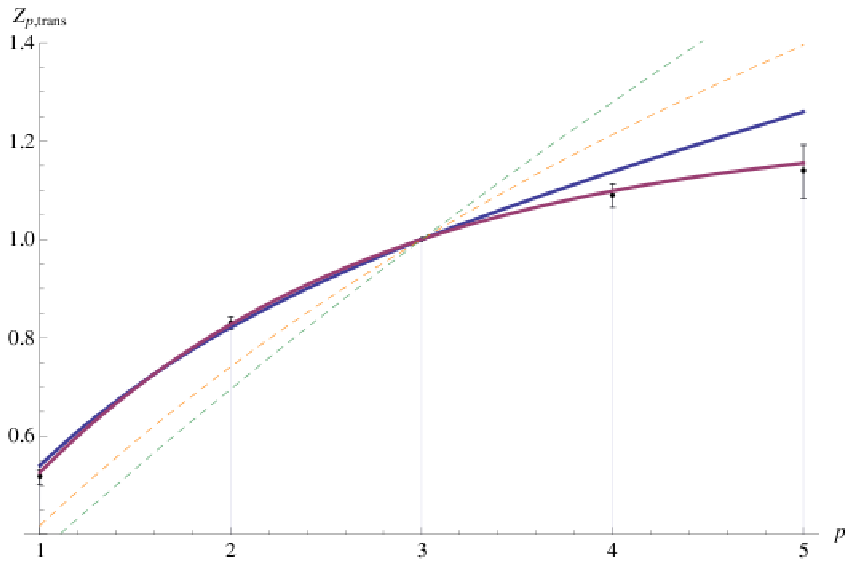}}
    \caption{Fits of the relative scaling exponents
    $Z_{p}=\zeta_{p}/\zeta_{3}$ to several models. The error bars
    indicate the scatter of the numerically calculated exponents, the
    dots specify the averages. The thin dashed lines show the scalings predicted by \citet{SheLeveque94} and by the Kolmogorov-Burger models~(\ref{eq:kolmogburg}) of \citet{Boldyrev02}, respectively. The general log-Poisson model~(\ref{eq:Z2par}) of \citet{Dub94} yields the two fits shown as thick solid lines. The fit function that closely matches the higher-order exponents corresponds to $\Delta=1$, whereas the other function was obtained with the constraint $\Delta=2/3$.}
    \label{pic:scaling}
  \end{center}
\end{figure}

For supersonic turbulence, in which dissipation is attributed to two-dimensional shocks, \citet{Boldyrev02} proposed to set $C=1$ while leaving the scaling index $\Delta=2/3$ unchanged. Then $\beta=1/3$
and, hence,
\begin{equation}
	\label{eq:kolmogburg}
	Z_{p} = \frac{p}{9}+1-\left(\frac{1}{3}\right)^{p/3}.
\end{equation}
The predictions of this so-called  Kolmogorov-Burgers model for $Z(1)$ and $Z_{2}$ agree very well with numerical results obtained from magnetohydrodynamic turbulence simulations with purely solenoidal forcing \citep{Boldyrev+02} and the supersonic turbulence simulation by \citet{KritNor07}. However, the time-averaged values of $Z_{p}$ obtained from our simulation differ significantly the Kolmogorov-Burgers model for all $p$ (see Table~\ref{tab:resultcomparison} and Figure~\ref{pic:scaling}). \citet{KritPad07} found deviations of the scaling exponents of mass-weighted velocity structure functions for $p>3$.

There are several factors, from which the discrepant scaling exponents might stem. Firstly,
it is possible that the computation of the structure functions is not well converged due
to insufficient sampling. Especially, this might affect the scaling exponents of higher order.
However, varying the sample size did not affect the obtained scaling exponents noticeably,
so we consider the calculated structure functions to be converged.
Secondly, the resolution of our simulation is possibly not high enough to allow for reliable estimates of the scaling exponents.  Other than the turbulence energy spectra, however, structure functions show a relatively broad power-law range since there is no bottleneck effect (see Table~\ref{tab:powerexponents} and Figure~\ref{pic:structurefunctions}) and ESS can be utilised to calculate the relative scaling exponents
even more reliably. Already from $256^{3}$ test runs, we were able to obtain estimates of the
relative scalings that showed the same trend as in the high-resolution simulation. Thirdly, we consider scaling exponents of transversal structure functions, while theoretical model usually refer to longitudinal structure functions. However, we found similar values of $Z_{1}$ and $Z_{2}$ from the longitudinal structure functions, while the disagreement of the higher-order scaling exponents with the Kolmogorov-Burgers model appears to be even more pronounced. Fourthly, the statistics might not be converged in time. This is excluded on grounds of the global statistics analysed at the beginning of Section~\ref{sc:results} and the small variations of the instantaneous scaling exponents listed in Table~\ref{tab:powerexponents}. There is possibly a small systematic drift of the early scaling exponents, but this has no significant influence on the results discussed in the following.

Given the relatively low RMS Mach number ($\mathcal{M}_\mathrm{rms}=2.2\ldots 2.5$) in our simulation, it is possible that vortices contribute more in comparison to shocks compared to the other simulations cited above, for which $\mathcal{M}_\mathrm{rms}>5$. On grounds of results from MHD simulations with different Mach numbers, \citet{PadJim04} proposed a one-parameter model based on equation~(\ref{eq:Z2par}), where $\beta=1-2/(3C)$ and the the codimension $C$ should monotonically decrease with $\mathcal{M}_\mathrm{rms}$ from $C=2$ in the subsonic limit to $C=1$ for sufficiently high RMS Mach numbers:
\begin{equation}
	\label{eq:Z1par}
	Z_{p} =\frac{p}{9}+C\left(1 - \left[1-\frac{2}{3C}\right]^{p/3}\right) .
\end{equation}
The best fit of this equation to the time-averaged scalings in our simulation yields $C\approx 0.71$, which falls outside the range of the one-parameter model. However, one should keep in mind that the corresponding fractal dimension $D\approx 2.3$ might very well result from the stretching and folding of dissipative structures in turbulent flow. Letting $p$ vary continuously, the corresponding curve is plotted in Figure~\ref{pic:scaling} together with the data points and other model predictions. 

If the notion of universal scaling in the sense of the one-parameter model was correct, then, particularly, the second-order exponent $Z_{2}$ would vary monotonically with the RMS Mach number form $0.7$ in the subsonic limit to $1.0$ in the hypersonic limit. However, \citet{KritNor07} obtained $Z_{2}\approx 0.76$, whereas $Z_{2}\approx 0.83$ for our simulation, although the RMS Mach number is more than twice as high in the former case. Thus, our findings point to non-universal scaling properties of supersonic turbulence. In addition to the RMS Mach number as major parameter, the intermittency of the compressible turbulence cascade might depend on the relative importance of compressive and solenoidal modes for the energy transfer among different scales. Indeed, the higher small-scale compressive ratio $r_{\mathrm{cs}}\approx 0.5$ in our simulation compared to about $0.3$ in the simulation of Kritsuk et al. indicates a larger fraction of dilatational (rotation-free) velocity fluctuations although the RMS Mach number is lower. Of course, this is a direct consequence of the applied forcing as far as large-scale velocity fluctuations are concerned. However, the divergence and the vorticity are mostly structural probes of small-scale velocity fluctuations \citep{Frisch}. For this reason, the result for $r_{\mathrm{cs}}$ suggests an influence of the forcing on the small-scale structure of turbulence. Remarkably, even in the incompressible limit, non-universal scaling at small scales has been found in numerical simulations of wall-bounded shear flow \citep{CasGual07} and for turbulence between counter-rotating cylinders \citep{AlexaMin05}. Particularly interesting is the spectral analysis presented in the latter work. It is shown that the turbulence energy flux at small scales is strongly influenced by triadic interactions with large scale modes and, thus, even distant scales cannot be separated. 

The one-parameter model~(\ref{eq:Z1par}) features the codimension $C$
as a measure of intermittency depending on the RMS Mach number, but
the scaling parameter $\Delta=2/3$ stems from the incompressible
turbulence cascade. However, it is not at all clear whether this
scaling applies to supersonic turbulence. Consequently, we fitted the
general two-parameter model~(\ref{eq:Z2par}) subject to the constraint
$\Delta\le 1$ to our data up to fourth order. The resulting parameters
are $\Delta\approx 1.00$ and $\beta\approx 0.57 < 2/3$. We also investigated the influence of including
$Z_{5}$ in the data set for the fit and found that the results did not change
significantly (the value of $C$ varied by less than $0.02$). However, releasing
the constraint $\Delta\le 1$, a value slightly greater than unity is
obtained for $\Delta$. Extending the fit to the scaling exponents up
to sixth oder, yields $\Delta\approx 1.05$. Therefore, higher-order
appears to be essential to constrain the $\Delta$-parameter. Remarkably, $\Delta=1$ follows from the scaling argument for the most singular dissipative structures that was used by \citet{SheLeveque94}, if it is assumed that the uniform dissipation time scale is proportional to $\ell$ as in Burgers turbulence.\footnote{The dissipation time scale is given by the kinetic energy $\sim u'^{2}(\ell)$ divided by the scale-invariant mean dissipation. For Burgers turbulence, $u'(\ell)\propto\ell^{1/2}$.} Therefore, the two-parameter fit implies shocks as the most singular dissipative structures. The $\beta$-parameter indicates a higher degree of intermittency in comparison to the Kolmogorov-Burgers model.  The corresponding codimension $C\approx 1.23$ is in accordance with the notion that the codimension varies form $C=2$ in the subsonic and $C=1$ in the hypersonic limit. 

\begin{figure}[t]
  \begin{center}
    \resizebox{\hsize}{!}{\includegraphics{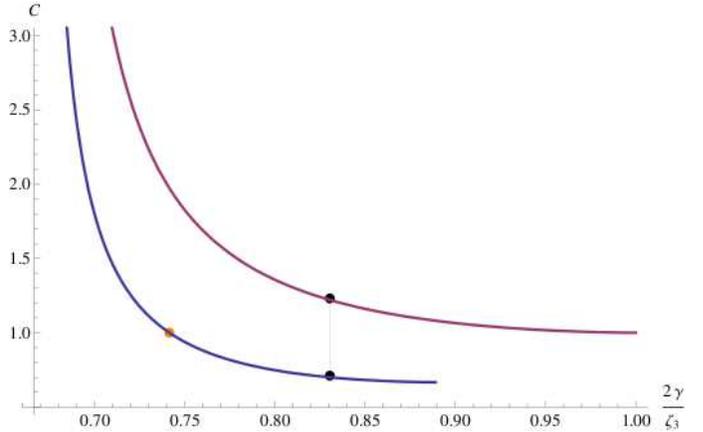}}
    \caption{Codimension $C=\Delta/(\beta-1)$ as function of the second-order scaling index $Z_{2}=2\gamma/\zeta_{3}$ for the log-Poisson models with $\Delta=2/3$ (lower line ) and $\Delta=1$
    (upper line), respectively. The two dots correspond to the fits to our numerical results shown in Figure~\ref{pic:scaling}. The Kolmogorov-Burgers model with $C=1$ is also marked by a dot.}
    \label{pic:codim}
  \end{center}
\end{figure}

Whereas both the one-parameter model with $C\approx 0.71$ and the
two-parameter model with $C\approx 1.23$ agree excellently with the
exponents $Z_{1}$ and $Z_{2}$, the latter matches the higher-order
statistics much better. However, the scaling exponents for $p>3$ are
subject to greater uncertainty (see Figure~\ref{pic:scaling}). For
this reason, we cannot  $C=0.71$. Interestingly, the same dichotomy with regard to $\Delta=2/3$ and $\Delta=1$ was found by \citet{KritPad07}, albeit for different values of $\beta$ and $C$. In this respect, it is interesting to investigate the dependence of the codimension $C$ on $Z_{2}$, which is related to the scaling index $\gamma$ via $Z_{2}=2\gamma/\zeta_{3}$. Substituting $\beta=1-\Delta/C$ and inverting equation~(\ref{eq:Z2par}) for $p=2$, the functions $C(2\gamma/\zeta_{3})$ for $\Delta=2/3$ and $\Delta=1$, respectively, are obtained. As one can see form the plots in Figure~\ref{pic:codim}, the solution for $\Delta=1$ converges towards $C=1$ as $Z_{2}\rightarrow 1$. Thus, the Burgers scaling follows from this model for $\gamma=0.5$ and $\zeta_{3}=1$. This corresponds to the interpretation that dissipation is completely dominated by shocks in the hypersonic limit. On the other hand, for $\Delta=2/3$, which includes the Kolmogorov-Burger model with $C(0.74)=2$, no real solution exists if $Z_{2}$ becomes greater than about $0.89$ (corresponding to $\gamma=0.45$). The minimal codimension is slightly less than $0.7$. Judging from the limiting behaviour of both models, it appears that the log-Poisson model with $\Delta=1$ is more sensible for supersonic turbulence.

Recently, measurements of higher-order scalings for diffuse molecular clouds in Polaris and Taurus have become available \citep{HilyFal08}. The relative scaling exponents obtained from these measurements, $Z_{2}\approx 0.69\ldots 0.71$, $Z_{4}\approx 1.27\ldots 1.30$ and $Z_{5}\approx 1.53\ldots 1.60$, fall in between the incompressible She-Leveque model and the Kolmogorov-Burgers model. The value $Z_{2}\approx 0.7$ can only be accommodated within the $\Delta=2/3$ branch, for which the codimension $C$ is close to $2$ (Figure~\ref{pic:codim}). This corresponds to weakly compressible, shear-dominated turbulence, plainly in opposition to the scenario discussed in this article. However, \citet{HilyFal08} consider only three different clouds with scaling exponents that are at there very lower end of those reported by \citep{BruntHey02b}. Consequently, a much broader sample of higher-order velocity statistics would be required in order to exclude the occurrence of compression-dominated turbulence in the ISM. 

\subsection{Density statistics}
\label{sc:discuss_dens}

\begin{figure}[t]
  \begin{center}
    \resizebox{\hsize}{!}{\includegraphics{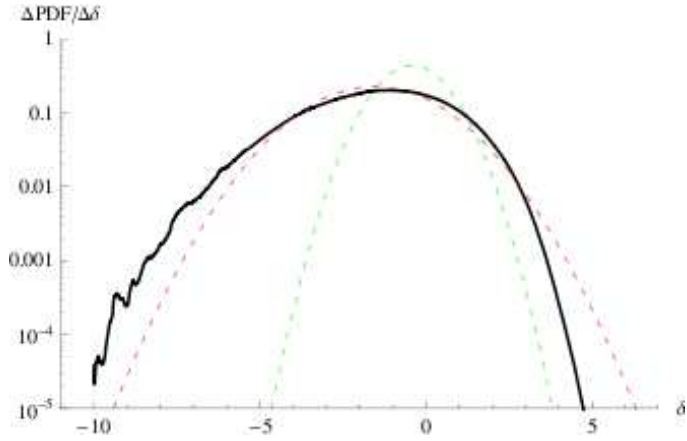}}
    \caption{Temporally averaged probability density function of the logarithmic mass density fluctuations, $\mathrm{ln}(\rho/\rho_{0})$, and log-normal fit functions (thin dashed lines).}
    \label{fg:avpdf}
  \end{center}
\end{figure}

Previous theoretical and numerical and studies favour log-normal statistics for the mass density of supersonic turbulence in isothermal gas \citep{Vaz94,PadNord97,PassVaz98,PadoanNordlund02}. This means that the distribution of $\delta=\ln(\rho/\rho_{0})$ is Gaussian. Consistency with $\langle\rho/\rho_{0}\rangle=1$ requires the variance of $\delta$ being related to the mean via $\sigma_{\delta}^{2}=-2\langle\delta\rangle$ . However, as one can see in Figure~\ref{fg:avpdf}, a normal distribution subject to this constraint does not properly fit the time-averaged $\delta$-pdf from our simulation. The standard deviation of the closest possible fit is $\sigma_{\mathrm{\delta}}\approx 1.76$. \citet{PadoanNordlund02} proposed that $\sigma_{\mathrm{\delta}}$ can be parameterised in terms of the RMS Mach number:
\begin{equation}
  \label{eq:padnord}
  \sigma_{\mathrm{\delta}}^{2}=\ln\left(1+0.25\mathcal{M}_{\mathrm{rms}}^{2}\right).
\end{equation}
We have $\mathcal{M}_{\mathrm{rms}}\approx 2.3$ in the interval of time over which the pdf of $\delta$ is averaged. For this Mach number, the above formula yields $\sigma_{\mathrm{\delta}}\approx 0.92$. The corresponding pdf is also plotted for comparison in Figure~\ref{fg:avpdf}. Clearly, compressively driven turbulence produces a  substantially broader range of density fluctuations than predicted by Padoan and Nordlund. \citet{KritPad07}, on the other hand, found a narrower range in
relation to the RMS Mach number.

\begin{figure*}[t]
  \begin{center}
    \mbox{\subfigure[]{\includegraphics[width=8.5cm]{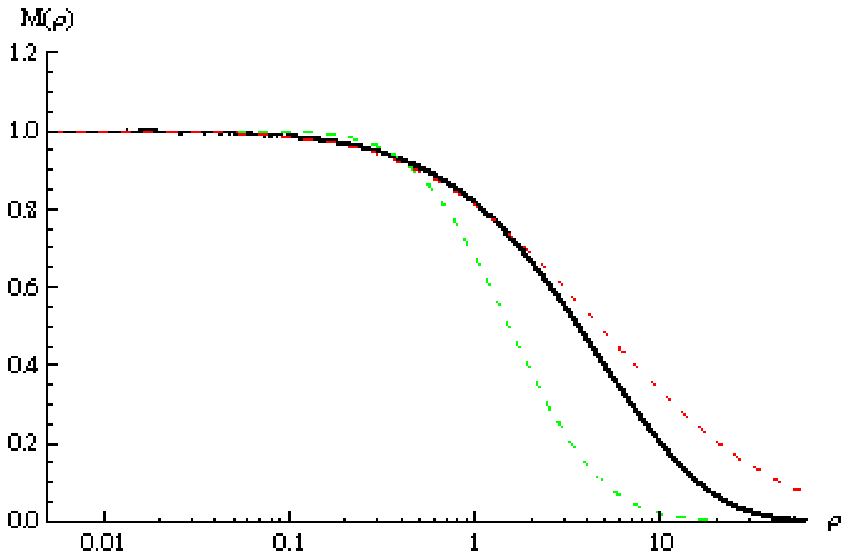}}\qquad
          \subfigure[]{\includegraphics[width=8.5cm]{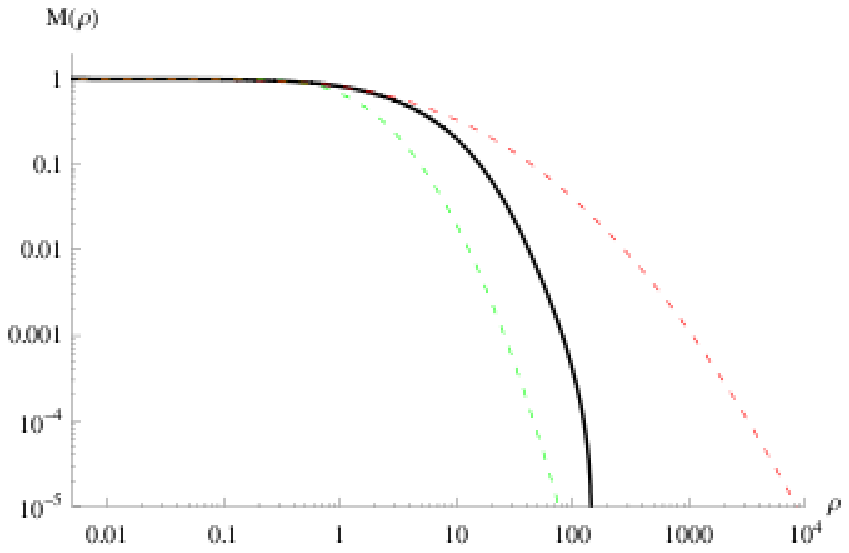}}}
   \caption{Integrated mass $M(\rho)$ in gas of density higher than $\rho$ in linear
   	scaling (a) and logarithmic scaling (b). The thick solid line results from the time-averaged
	pdf obtained from our simulation. The closest log-normal distribution and the log-normal
	distribution implied by the Padoan-Nordlund relation~\ref{eq:padnord} yield the dashed cures.
	While the former greatly overestimates the mass in the density peaks, the latter underestimates 
	the fraction of over-dense gas.}
    \label{fg:intgr_mass}
  \end{center}
\end{figure*}

Since we have sampled the $\delta$-pdfs over many integral time scales, the shape of our time-averaged pdf is very likely genuine. In test simulations, we have also excluded the possibilities that it is purely a resolution effect or a consequence of the nearly isothermal approximation. A skewed distribution of density fluctuations has important consequences. In Figure~\ref{fg:intgr_mass}, the total mass of gas with density higher than a given threshold density,
\begin{equation}
	M(\rho) = 
	(2L)^{3}\int_{\rho}^{\infty}\rho' \mathrm{pdf}(\rho')\,\dd\rho' =
	\rho_{0}(2L)^{3}\int_{\delta}^{\infty}\exp(\delta)\,\mathrm{pdf}(\delta)\,\dd\delta',
\end{equation}
is plotted for the time-averaged probability density function $\mathrm{pdf}(\delta)$ from our simulation,
closest log-normal pdf and the log-normal pdf with $\sigma_{\mathrm{\delta}}\approx 0.92$, respectively. It is palpable that calculating $M(\rho)$ on the basis of the log-normal fit with $\sigma_{\mathrm{\delta}}\approx 1.76$ in place of the numerical pdf implies enormously wrong mass fractions in the high-density peaks. On other hand, the prediction of $M(\rho)$ based on the Padoan-Nordlund relation~(\ref{eq:padnord}) underestimates the total mass of highly compressed gas by about one order of magnitude.  For $\rho\gtrsim 100$, a numerical cutoff can be discerned, as the mass function corresponding to the numerical pdf plunges towards zero in the logarithmically scaled plot (b). The reason is that density fluctuations cannot become arbitrarily high due to the
spatial discretisation and, because of the discretisation in time, the most intermittent events are too rare to show up in the statistics. The mass range affected by the numerical cutoff is $M(\rho)\lesssim 10^{-3}$. We found that this range largely overlaps with the mass range for the computation of clump mass spectra following the prescription by \citet{PadNord07}.
The high sensitivity of mass spectra on numerical resolution has already been noted by \cite{HenneAud07}. Consequently, it is mandatory to go to substantially higher resolutions which, in turn, necessitates the application of adaptive mesh refinement. Unlike the AMR simulation at hand, extreme density peaks have to be followed to very high resolution
and, at the same time, turbulent flow has to be treated self-consistently in less resolved regions of lower gas density.

A justification for log-normal pdfs of the mass density based on analytical arguments was given by \citet{PassVaz98} for one-dimensional isothermal gas dynamics without external forces. These arguments
also apply to fully developed turbulence in the strict sense, i.~e., three-dimensional turbulence that evolves without imposed constraints such as boundaries, external forces or viscosity. To what extent
can this be carried over to turbulence in a periodic box with stochastic forcing? Setting the pressure $P=c_{0}^2\rho$, where $c_{0}$ is the isothermal speed of sound, the pressure gradient divided by the mass density is given by
\begin{equation}
	\frac{1}{\rho}\vect{\nabla}P =
	c_{0}^{2}\vect{\nabla}\delta.
\end{equation}
Thus, the balance laws~(\ref{eq:mass}) and~(\ref{eq:vel}) can be written in the form
\begin{align}
  \label{eq:mass_isoth}
  \frac{\DD}{\DD t}\delta &= -d, \\
  \label{eq:vel_isoth}
  \frac{\DD}{\DD t}\vect{u} &=
  -c_{0}^{2}\vect{\nabla}\delta + \vect{f}.
\end{align}
The first equation means that infinitesimal Lagrangian changes of the density are given by $-d\,\dd t$.
It is argued that density fluctuations are built up in a hierarchical process, meaning that $\delta$ evolves as a random process, where infinitesimal increments $\delta_{+}\dd t$ (density enhancements) and decrements $\delta_{-}\dd t$ (density reductions) are added with equal probability independent of the value of $\delta$. The central limit theorem then implies a Gaussian distribution of $\delta$. This is the just the Wiener process that was mentioned in Section~\ref{sc:numerics}. These arguments also apply to the three-dimensional case. However, the action of an external force can cause deviations from the log-normal statistics. Let us consider equation~(\ref{eq:div_isoth}) for the rate of compression in the isothermal case. In regions, in which $\vect{\nabla}\cdot\vect{f}<0$, the force tends to increase the rate of compression. The equation for the density fluctuations~(\ref{eq:mass_isoth}) implies that the corresponding fractional change of $-d\,\dd t$ results in a stronger increment $\delta_{+}\dd t$ or a smaller decrement $\delta_{-}\dd t$, i.~e., a force with positive convergence locally supports compression and weakens rarefaction, as one would expect. In the case $\vect{\nabla}\cdot\vect{f}>0$, the force has the opposite effect. If the force field is stochastic, both effects will occur with equal probability at any time at any position (by the very construction of the force field). Hence, the net effect depends on the periods of time a particular fluid parcel is contracting ($-d>0$) or expanding ($-d<0$). Apart from that, the effect will weaken toward smaller scales and higher densities (because of scaling considerations). For a detailed analysis, Lagrangian density statistics would be required. From the existing simulation, we have only Eulerian statistics. Nevertheless, our considerations show that, in general, forces with a strong dilatational component may cause deviations form a log-normal distribution of the density fluctuations in isothermal gas. This is what we observe in our simulation. 

The spectral properties of $\delta$ in our simulation indicate an intermittent self-similar structure,
where the scaling appears to be closely linked to the scaling of the velocity fluctuations. The
spectral index $\beta_\delta\approx 1.84$ implies an exponent $\gamma_{\delta}\approx 0.42$
of the RMS logarithmic density fluctuation. The corresponding index of the RMS velocity fluctuations
is $\gamma\approx 0.43$. Similar scaling properties of the velocity and density fluctuations are
implied by equation~(\ref{eq:mass_isoth}) provided that $\vect{f}$ only acts on large scales. Therefore, self-similarity (related to scaling) appears to be a more robust property of turbulent density fluctuations than log-normality (a probabilistic concept), contrary to the explanation of log-normal statistics on grounds of self-similarity in the literature.

\section{Conclusion}

We performed simulations of supersonic turbulence in isothermal gas driven by a mostly compressive stochastic force field. The analysis of statistical properties of the velocity and mass density revealed several remarkable results:

\begin{enumerate}
\item The turbulence energy spectrum function is close to the spectrum of Burgers turbulence
in the initial phase of the flow evolution and subsequently approaches a flatter spectrum with
an index $\beta\approx 1.9$. Within the error bars, a comparable index $\beta_{\delta}\approx 1.8$ is found for the spectrum function of $\delta^{2}$, where $\delta=\ln(\rho/\rho_{0})$ is the logarithmic density fluctuation.
\item In the statistically stationary state, the scaling exponent of the RMS velocity fluctuation $\gamma=\zeta_{2}/2\approx 0.43$, which is comparable to typical exponents inferred from measurements of the velocity dispersion in molecular clouds.
\item The relative scaling exponents $Z_{p}=\zeta_{p}/\zeta_{3}$ up to order $p=5$ deviate from the Kolmogorov-Burgers model for supersonic turbulence. Varying only the codimension $C$ of dissipative structures, the best fit to our data implies $C\approx 0.71$, whereas $C=1$ in the Kolmogorov-Burgers model. Fitting the general log-Poisson hierarchical structure model, we find $C\approx 1.23$ and the scaling index $\Delta\approx 1.0$. This can be interpreted as shocks being the \emph{most singular} dissipative structures. In the Kolmogorov-Burgers model, $\Delta=2/3$ is
assumed.
\item Compared to other simulations, our analysis point toward non-universal scaling properties, i.~e., the velocity statistics cannot be accommodated within a one-parameter family of models. There appears to be an additional degree of freedom related to the large-scale forcing.
\item The time-averaged probability density function of $\delta$ is not consistent with a log-normal distribution of the mass density. We find a skewed distribution that is broader than
the log-normal distribution predicted by \citet{PadoanNordlund02} on the basis of the RMS Mach number. Large deviations of the mass fraction in over-dense gas compared to log-normal distributions are implied.
\end{enumerate}

The assumption of a log-normal mass density distribution has been used for analytical predictions
of the rate of turbulence-regulated star formation \citep{PadoanNordlund02,KrumKee05,HenneChab08}. With the distribution of the mass density found in our simulation, the resulting star formation rate would differ substantially form the log-normal case because the mass converted into stars by gravitational collapse mainly depends on the high-density tail of the probability distribution function. However, this result is not conclusive yet, because a wider range of parameters has to be studied, self-gravity and magnetic fields have to be included, and, most importantly, the dependence of the mass density distribution and clump mass spectra on the numerical resolution has to be clarified. In part, this is the subject of ongoing work, where turbulence simulations with purely solenoidal and compressive forcing are systematically compared for numerical resolutions up to $1024^{3}$ \citep{SchmFeder08,FederKless08}. Moreover, this complementary study is essential to corroborate the non-universal velocity scaling properties discussed in this article.

The observational results of \citet{BruntHey02b} point towards a scaling exponent $\alpha$ that is \emph{significantly greater} than $0.5$ in many instances. This is hard to explain on the basis of supersonic turbulence in isothermal gas. Stiffer scaling exponents might be caused, for instance,
by self-gravity \citep{BigDia89}. On the other hand, \citet{HilyFal08} found scaling properties that are indicative of nearly incompressible, shear-dominated turbulence in some molecular clouds. Possibly, molecular clouds are a rather inhomogeneous class of objects. From that point of view, the situation is reminiscent of the observation of type Ia supernovae, which were recognised to be much more diverse than it had appeared in the beginning. To understand how the mass density distribution and the two-point statistics of the turbulent velocity at various scales in the interstellar medium comes about, it is vital to include additional physics such as self-gravity, radiative cooling and magnetic fields also in numerical simulations of turbulence with stochastic forcing. To this end, the application of adaptive methods will be indispensable. \citet{OffKrum07} succeeded in performing simulations of self-gravitating isothermal turbulence with AMR. Radiative post-processing of their numerical data revealed discrepancies with observed core line widths both for driven and for decaying turbulence. This suggests that not only gravity but also thermal processes are important for the dynamics of molecular cloud turbulence. With the AMR simulation presented in this article, we have attempted a first step toward a methodology that is based on the dynamics of turbulent flows. \citet{IapiAdam08} have already presented a multi-level AMR application using these techniques. The development is far from being complete yet, but it appears to be promising for a self-consistent treatment of turbulence in the complex scenarios encountered in star-forming clouds.

\begin{acknowledgements}

We thank Alexei Kritsuk and Paolo Padoan for numerous discussions at KITP, Santa Barbara,
which helped to enrich this article.  We also thank Emmanuel Leveque for his comments on the
intermittency models. Special thanks to Leonhard Scheck for rendering the 3D visualizations. The simulations were performed in the framework of the DEISA Extreme Computing Initiative on SARA Aster, Netherlands, and HLRB2, Leibniz Supercomputing Centre in Garching. W. Schmidt was supported by the Elite Network of Bavaria. C. Federrath is fellow of the International Max Planck Research School for Astronomy and Cosmic Physics at the University of Heidelberg.

\end{acknowledgements}

\bibliographystyle{aa}
\bibliography{ComprTurb_I}

\end{document}